\documentclass[twocolumn]{autart}

\usepackage{graphicx}
\usepackage{bm}

\usepackage{indentfirst}

\usepackage{booktabs} 
\usepackage{diagbox}
\usepackage{amsmath}

\usepackage{amssymb}
\usepackage{amsfonts}
\usepackage{algorithm}
\usepackage{algpseudocode}
\usepackage{mathrsfs}
\usepackage{bbm}

\usepackage{color,xcolor}

\usepackage{epsfig}

\ifx\pdfoutput\undefined 
\usepackage{graphicx}
\DeclareGraphicsExtensions{.eps,.ps} \else
\usepackage{graphicx}
\DeclareGraphicsExtensions{.pdf} \fi
\newtheorem{remark}{Remark}
\begin{document}
	
\begin{frontmatter}
	
\title{Multiple-Source Ellipsoidal Localization Using Acoustic Energy Measurements\thanksref{footnoteinfo}}

\thanks[footnoteinfo]{This work was supported in part by the NSFC No. 61673282, U1836103 and the PCSIRT16R53.}

\author[A,B]{Fanqin Meng}\ead{mengfanqin2008@163.com},
\author[C]{Xiaojing Shen}\ead{shenxj@scu.edu.cn},
\author[D,D2]{Zhiguo Wang}\ead{wangzg315@126.com},
\author[C]{Haiqi Liu}\ead{haiqiliu0330@163.com},
\author[A]{Junfeng Wang}\ead{wangjf@scu.edu.cn},
\author[C]{Yunmin Zhu}\ead{ymzhu@scu.edu.cn},
\author[E]{Pramod K. Varshney}\ead{varshney@syr.edu}

\address[A]{School of Aeronautics  $\&$ Astronautics, Sichuan University, Chengdu, Sichuan 610064, China}
\address[B]{School of Automation $\&$ Information Engineering,  Sichuan University of Science $\&$ Engineering, Yibin, Sichuan 644000, China}
\address[C]{School of Mathematics, Sichuan University, Chengdu, Sichuan 610064, China}
\address[D]{School of Science $\&$ Engineering, The Chinese University of Hong Kong, Shenzhen, Guangdong 518172,  China}
\address[D2]{Department of Electronic Engineering and Information Science, University of Science and Technology of China, Hefei, Anhui 230026, China.}
\address[E]{Department of Electrical Engineering  $\&$ Computer Science, Syracuse University, NY 13244, USA}


\begin{abstract}

In this paper, the  multiple-source ellipsoidal localization problem based on acoustic energy measurements is investigated via  set-membership estimation theory. When the probability density function of measurement noise is unknown-but-bounded, multiple-source localization is a difficult problem since not only the acoustic energy measurements are complicated nonlinear functions of multiple sources, but also the multiple sources bring about a high-dimensional state estimation problem.  First,  when the  energy parameter  and the position of the source are bounded in an interval and a ball  respectively, the nonlinear remainder bound of the Taylor series  expansion   is  obtained analytically on-line. Next, based on the separability of the nonlinear measurement function, an efficient estimation procedure is developed. It solves  the multiple-source localization problem  by using an alternating optimization iterative algorithm, in which the remainder bound needs to be known on-line. For this  reason,  we first derive  the remainder bound analytically. 
When the energy decay factor is unknown but bounded, an efficient estimation procedure is developed based on interval mathematics. 
Finally, numerical examples  demonstrate the effectiveness of the ellipsoidal localization algorithms  for multiple-source localization. In particular, our results  show that when the noise is non-Gaussian, the set-membership localization algorithm performs better than the EM  localization algorithm.
\end{abstract}

\end{frontmatter}

\section{Introduction}\label{sec_1}

Localization is an important research problem in many systems such as radar, sonar and multimedia systems. Source localization using a network of sensors has far-reaching applications, e.g., battlefield security, surveillance, environment or health monitoring and disaster relief operations. Many works have  investigated the single-source localization problem (see \cite{Bishop-Fidan-Anderson-Dogancay-Pathirana10},  \cite{Chen-Dai-Shen-Lau-Win16},
\cite{Han-Kieffer-Lambert17}, \cite{Masazade-Niu-Varshney-Keskinoz10}, \cite{Niu-Blum-Varshney-Drozd12}). However, very limited work has  been reported on  the multiple-source localization problem. In this paper, we focus on the multiple-source localization problem where the aim is to estimate the coordinates of multiple  acoustic sources.

The problem of source localization has been addressed by many authors (see papers  \cite{Bishop-Fidan-Anderson-Dogancay-Pathirana10}, \cite{Chen-Dai-Shen-Lau-Win16}, \cite{Han-Kieffer-Lambert17},   
 \cite{Masazade-Niu-Varshney-Keskinoz10}, \cite{Niu-Blum-Varshney-Drozd12}, \cite{Niu-Vempaty-Varshney18}, \cite{Piovana-Shames-Fidanc-Bulloa-Andersond13},  \cite{Shames-Fidan-Anderson09},  \cite{Vempaty-Han-Varshney14}, \cite{Vempaty-Ozdemir-Agrawal-Chen-Varshney13}, 
  \cite{Win-Dai-Shen-Chrisikos-Poor18}, \cite{Wymeersch-Lien-Win09}
and books  \cite{Mao-Fidan09}, \cite{Strumillo11}, \cite{Reza-Buehrer11}).
Most localization methods are based on one of the following three types of physical variables measured by sensor readings for localization: direction of arrival (DOA), time difference of arrival (TDOA) and received sensor signal strength (RSS). 
DOA can be estimated by exploiting the phase difference measured at receiving sensors (see  \cite{Wu-Wong12}, \cite{Zhang-Xu-Xu10}).  TDOA is based upon the difference in arrival times of the emitted signals received at a pair of sensors (see  \cite{Fowler-Hu08}, \cite{Malanowski-Kulpa12}, \cite{Mellen-Pachter-Raquet03}). The  source localization estimation task  with DOA and TDOA can be performed by solving a nonlinear least squares (NLS) problem. These methods mainly deal with the single target localization problem.

For the multiple-source localization problem, the maximum-likelihood (ML) method  is widely used (see  \cite{Chen-Hudson-Yao02},  \cite{Meng-Xiao17}, \cite{Sheng-Hu05}, \cite{Wang-Quost-Chazot-Antoni16}). A multiresolution  search  and the expectation maximization (EM) method \cite{Sheng-Hu05} were proposed to solve the multiple-source localization problem. An efficient EM algorithm \cite{Meng-Xiao-Xie11} was proposed to improve  estimation accuracy.  Authors use the model which is called the acoustic energy decay model based on RSS to solve the multiple-source localization problem. The source locations and strengths are estimated using a variant of the EM algorithm in \cite{Wang-Quost-Chazot-Antoni16} with Helmholtz operator. In this paper, we focus on the acoustic energy decay model mentioned in \cite{Chen-Hudson-Yao02},  \cite{Meng-Xiao17}, \cite{Sheng-Hu05}. 
The measurement noise is modeled as  additive white Gaussian noise in these articles.
When the unknown noise is not Gaussian, this approach may lead to poor performance because it  is sensitive to the exact probabilistic knowledge of the parameters of  noise (see \cite{Theodor-Shaked-deSouza94}).
In practice,  the assumed probabilistic model may not be accurate resulting in model mismatch.
It then seems more natural to assume that the state perturbations and
measurement noise are unknown but bounded
(see \cite{Polyak-Nazin-Durieu-Walter04}). Under these assumptions, the articles \cite{Caiti-Garulli-Livide-Prattichizzo05}, \cite{Garulli-Vicino01}, \cite{Jaulin09} and \cite{Yu-Zamora-Soria16} discussed  the single source localization problem for different applications. However,	they do not consider the multiple source localization problem with acoustic energy decay model.
These facts motivate us to further research the multiple-source ellipsoidal localization problem under  the unknown-but-bounded measurement noise assumption.

When the measurement noise is unknown-but-bounded, 
set-membership estimation theory may be used to solve the
multiple-source localization problem since it does not require
a statistical description of the errors. Set-membership
estimation was considered first in  1960s (see \cite{Bertsekas-Rhodes71}, \cite{Schweppe68}, \cite{Witsenhausen68}).
The critical step  here  is the computation of  bounding ellipsoids (or boxes,
simplexes, parallelotopes, and polytopes) which are guaranteed
to contain the state vector to be estimated given bounds on the
perturbations and noises. The problem of bounding ellipsoids has been extensively investigated,
for example, see papers \cite{Calafiore05}, \cite{Durieu-Walter-Polyak01},
\cite{Shen-Zhu-Song-Luo11}, the book \cite{Jaulin-Kieffer-Didrit-Walter01}, and
references therein. However, the ellipsoidal bounding method has not been investigated for the solution of  the multiple-source
localization problem using acoustic energy measurements.

In this paper, we attempt to make progress on the multiple-source localization problem based on acoustic energy measurements in the bounded noise setting by the ellipsoidal bounding estimation method. Multiple-source  localization is a difficult problem. There are two main difficulties: the acoustic energy measurements are complicated nonlinear functions of multiple sources and the multiple sources lead to a  high-dimensional state estimation problem. The main contributions of this paper are as follows. First, when the parameter is bounded in a convex set, the remainder bound is obtained by taking samples on the boundary of the set.
Moreover,  when the energy parameter and  the position of the source are bounded in  an interval and a ball respectively, the remainder bound can be obtained  analytically on-line.
Next, an efficient procedure is developed to solve   the multiple-source localization problem  using an alternating optimization iterative algorithm. Furthermore, an efficient estimation procedure is developed based on interval mathematics when the energy decay factor is unknown but bounded.  Numerical examples show that when the measurement noise is unknown-but-bounded, the performance of the ellipsoidal localization algorithm is better than that of the EM localization algorithm. Some preliminary results on this problem were presented at a conference \cite{Meng-Shen-Wang-Zhu17}. This paper now includes all the mathematical details and proofs. 

The rest of this paper  is organized as follows. Preliminaries are given in Section \ref{sec_2}.
In Section \ref{sec_3}, the bounding set of the remainder  is obtained from the boundary of the state bounding ellipsoid. Moreover, the bounding set is obtained analytically  when the energy parameter and the position of the source are bounded
in an interval and a ball  respectively.
In Section \ref{sec_4}, the solution to the multiple-source ellipsoidal localization problem  is derived by solving an SDP problem based on S-procedure and Schur complement. In Section \ref{sect-1}, an interval mathematics  estimation method is developed to deal with the multiple-source localization problem when the  energy decay factor is unknown but bounded. In Section \ref{sec_6}, numerical examples are given and discussed. Finally, Section \ref{sec_7} is devoted to concluding remarks.

\section{Preliminaries}\label{sec_2}
\subsection{Acoustic Energy Attenuation Model}

The acoustic energy attenuation model is adopted in this paper (see, e.g., \cite{Sheng-Hu05}). Consider a sensor network composed of $L$ sensors distributed at known spatial locations, denoted $r_l,\ l=1,\cdots,L$, where $r_l\in \mathbb{R}^d$,  $d=2$ or $3$. A fusion center is used to collect the measurement data of the sensors and to run the source localization algorithm. There are $N$  acoustic sources whose locations need to be determined. The number of sources $N$ is  known. The sources are  static and the locations of the sources are denoted by $\rho_n\in \mathbb{R}^d,\ n=1,\cdots,N$, which are unknown. 
Each sensor considers only a single RSS measurement and it is expressed as
\begin{align}\label{eqsm_2}
y_l=g_l\sum_{n=1}^{N}\frac{s_{n}}{\parallel\rho_n-r_l\parallel^{\alpha}}+\varepsilon_l,\ l=1,\cdots,L,
\end{align}
where $s_n$ is a scalar denoting the energy emitted by the $n$-th source,  $\parallel\rho_n-r_l\parallel\neq0$ is the distance between the $n$-th source and the $l$-th sensor, 
$g_l$ is the gain factor of the $l$-th acoustic sensor, $\alpha$ is a known energy decay factor with a typical value lies between  2 to 4 (see \cite{Meng-Xiao-Xie11}), and the additive measurement noises  $\varepsilon_l,\ l=1,\ldots,L$, are  independent and unknown-but-bounded, i.e., $\bm{\varepsilon}=[\varepsilon_1,\varepsilon_2,\cdots,\varepsilon_L]^T$ is confined to a specified box 
\begin{align}\label{eqsm_5}
\begin{split}
\mathcal{B}^{\varepsilon}&=\{\bm{\varepsilon}\in \mathbb{R}^L:\bm{D}^{\varepsilon}_l\leq\varepsilon_l\leq\bm{U}^{\varepsilon}_l,l=1,\cdots,L\}, 
\end{split}
\end{align}
where $\bm{D}^{\varepsilon}_l$ is the $l$-th component of the lower bound $\bm{D}^{\varepsilon}$ of the box $\mathcal{B}^{\varepsilon}$, i.e., $\bm{D}^{\varepsilon}=[\bm{D}^{\varepsilon}_1,\cdots,\bm{D}^{\varepsilon}_L]^T$, $\bm{U}^{\varepsilon}_l$ is the $l$-th component of the upper bound $\bm{U}^{\varepsilon}$ of the box $\mathcal{B}^{\varepsilon}$, i.e., $\bm{U}^{\varepsilon}=[\bm{U}^{\varepsilon}_1,\cdots,\bm{U}^{\varepsilon}_L]^T$, and $[\cdot]^T$ denotes the transpose of $[\cdot]$.

Moreover,  the scalar $s_n$ is independent of  the position $\rho_n$ of the $n$-th source  and the unknown parameters  of the different sources are independent.
The unknown parameters  of the $n$-th source are $s_n$ and $\rho_n$, which are denoted as $\bm{x}_n$, i.e., $\bm{x}_n=[s_n,\rho_n^T]^T$, $n=1,\cdots, N$. All the unknown parameters  of the $N$ sources are concatenated  and denoted as $\bm{x}=[\bm{x}_1^T,\bm{x}_2^T,\cdots,\bm{x}_N^T]^T$. All the measurements of the $L$ sensors are  denoted as $\bm{y}=[y_1,y_2,\cdots,y_L]^T$.
We define the following notation:
\begin{align}\label{mfq_1}
f(\bm{x})&=[\sum_{n=1}^Nf_{n,1}(\bm{x}_n),\cdots,\sum_{n=1}^Nf_{n,L}(\bm{x}_n)]^T,
\end{align}
where 
$f_{n,l}(\bm{x}_n)=g_l\frac{s_n}{\parallel\rho_n-r_l\parallel^{\alpha}}$.
The acoustic energy measurement functions (\ref{eqsm_2}) used  for  multi-source localization are written in a simpler notation as
\begin{align}\label{eqsm_3}
\bm{y}=f(\bm{x})+\bm{\varepsilon},
\end{align}
where $\bm{\varepsilon}=[\varepsilon_1,\varepsilon_2,\cdots,\varepsilon_L]^T$ is the additive measurement noise.

\subsection{Multiple-source Ellipsoidal Localization Problem}
The bounding set $\mathcal{E}$ of the state $\bm{x}$ of the $N$ sources is considered as the Cartesian product of $\mathcal{E}_n$, i.e.,
\begin{align} \label{eqsm_14_2}
\begin{split}
\mathcal{E}&=\prod_{n=1}^N\mathcal{E}_n,
\end{split}
\end{align}
where $\mathcal{E}_n$ is the bounding set of the state $\bm{x}_n$. Since the scalar $s_n$ is independent of the position $\rho_n$ of the $n$-th source,  $s_n$ is  contained in an interval $\mathcal{E}_{n}^s=\{s\in\mathbb{R}^{1}: |s-\hat{s}_{n}|\leq S_{n}\}$ and $\rho_n$ is  contained in an ellipsoid
\begin{align*}
\mathcal{E}_{n}^{\rho}=\{\rho\in\mathbb{R}^{d}: (\rho-\hat{\rho}_{n})^T \bm{P}_{n}^{-1}(\rho-\hat{\rho}_{n})\leq1\},
\end{align*}
where $\hat{\rho}_{n}$  is the center of the ellipsoid $\mathcal{E}_{n}^{\rho}$, and $\bm{P}_{n}$ is the $shape\ matrix$ of the ellipsoid $\mathcal{E}_{n}^{\rho}$.
Then the bounding set $\mathcal{E}_n$ is
\begin{align} \label{eqsm_14}
\begin{split}
\mathcal{E}_n&=\mathcal{E}_{n}^s\times\mathcal{E}_{n}^{\rho}.
\end{split}
\end{align}

When the nonlinear measurement function $f$ is linearized, the remainder term is bounded by a box. Specifically, by Taylor's Theorem, $f$  is linearized to
\begin{align}\label{eqsm_11}
\begin{split}
f(\bm{x})=f(\hat{\bm{x}})+&\bm{J}(\hat{\bm{x}})(\bm{x}-\hat{\bm{x}})+\Delta f(\bm{x},\hat{\bm{x}}),
\end{split}
\end{align}
where $\hat{\bm{x}}=[\hat{\bm{x}}_1^T,\cdots,\hat{\bm{x}}_N^T]^T$, $\hat{\bm{x}}_n=[\hat{s}_n,\hat{\rho}_n^T]^T$, $n=1,\cdots,N$,  $\bm{J}(\hat{\bm{x}})=\frac{\partial f(\bm{x})}{\partial \bm{x}}|_{\hat{\bm{x}}}$ is the Jacobian matrix, and $\Delta f(\bm{x},\hat{\bm{x}})$ is the higher-order remainder which is bounded in a box $\mathcal{B}$ for all $\bm{x}\in\mathcal{E}$, i.e.,
\begin{align}\label{eqsm_12}
\begin{split}
\Delta f(\bm{x},\hat{\bm{x}})\in\mathcal{B}=&\{\bm{z}\in \mathbb{R}^L:\bm{D}_l^f\leq \bm{z}_l\leq \bm{U}_l^f,\\
&~~~~~~~~~~~~\ l=1,\cdots,L\},
\end{split}
\end{align}
where $\bm{D}^{f}_l$ is the $l$-th component of the lower bound $\bm{D}^{f}$ of the box $\mathcal{B}$, i.e., $\bm{D}^{f}=[\bm{D}^{f}_1,\cdots,\bm{D}^{f}_L]^T$, $\bm{U}^{f}_l$ is the $l$-th component of the upper bound $\bm{U}^{f}$ of the box $\mathcal{B}$, i.e., $\bm{U}^{f}=[\bm{U}^{f}_1,\cdots,\bm{U}^{f}_L]^T$. 
Note that we do not assume that the box $\mathcal{B}$ is given before the algorithm. It is determined  on-line.

We consider  an efficient estimation procedure  to solve the multiple-source localization problem by using
an alternating optimization iterative approach. It is formulated as follows.   Assume that the state $\bm{x}$ belongs to a given initial bounding set $\mathcal{E}^0$, which is the Cartesian product of  ellipsoids $\mathcal{E}_{n}^0, n=1,\ldots,N$, i.e.,
\begin{align}\label{eqsm_8}
\begin{split}
\mathcal{E}^0&=\prod_{n=1}^N\mathcal{E}_{n}^0=\prod_{n=1}^N\mathcal{E}_{n}^{s,0}\times\mathcal{E}_{n}^{\rho,0},
\end{split}
\end{align}
where $\mathcal{E}_{n}^{s,0}=\{s\in\mathbb{R}^{1}: |s-\hat{s}_{n}^0|\leq S_{n}^0\}$, and $\mathcal{E}_{n}^{\rho,0}=\{\rho\in\mathbb{R}^{d}: (\rho-\hat{\rho}_{n}^0)^T (\bm{P}_{n}^0)^{-1}(\rho-\hat{\rho}_{n}^0)\leq1\}$.

At the $i$-th iteration, given that $\bm{x}$ belongs to the current bounding set $\mathcal{E}^i,$
\begin{align}\label{eqsm_9}
\begin{split}
\mathcal{E}^i&=\prod_{n=1}^N\mathcal{E}_{n}^i=\prod_{n=1}^N\mathcal{E}_{n}^{s,i}\times\mathcal{E}_{n}^{\rho,i},
\end{split}
\end{align}
where $\mathcal{E}_{n}^{s,i}=\{s\in\mathbb{R}^{1}: |s-\hat{s}_{n}^i|\leq S_{n}^i\}$, and $\mathcal{E}_{n}^{\rho,i}=\{\rho\in\mathbb{R}^{d}: (\rho-\hat{\rho}_{n}^i)^T (\bm{P}_{n}^i)^{-1}(\rho-\hat{\rho}_{n}^i)\leq1\}$.

At the ($i+1$)-th iteration, based on the measurement $\bm{y}$, the goal of the ellipsoidal localization estimation algorithm  is to determine a bounding set $\mathcal{E}^{i+1}=\Pi_{n=1}^N\mathcal{E}_{n}^{i+1}$, whenever \uppercase\expandafter{\romannumeral1}) $\bm{x}$ is in $\mathcal{E}^{i}$, \uppercase\expandafter{\romannumeral2}) the measurement noise $\bm{\varepsilon}\in\mathcal{B}^{\varepsilon}$ and the remainder $\Delta f(\bm{x},\hat{\bm{x}}^i)\in\mathcal{B}^i$.

Moreover,  the shape matrix of the state bounding set $\mathcal{E}^i$ is denoted as $\bm{P}^i$ and 
\begin{align*}
\bm{P}^i=\begin{bmatrix}
 \bm{P}_{1}^i&0&0&0 &\cdots& 0&0\\
 0&s_{1}^i&0&0&\cdots&0&0\\
 0&0&\bm{P}_{2}^i&0&\cdots&0&0\\
  0&0&0&s_{2}^i&\cdots&0&0\\
  \vdots&\vdots&\vdots&\vdots&\vdots&\vdots&\vdots\\
  0&0&0&0&\cdots&\bm{P}_{N}^i&0\\
  0&0&0&0&\cdots&0&s_{N}^i
\end{bmatrix}.
\end{align*}
We provide a state bounding set $\mathcal{E}^i$ by minimizing its ``size" at $i$-th iteration which is a function of the shape matrix $\bm{P}^i$ denoted by $g(\bm{P}^i)$. Throughout this paper, $g(\bm{P}^i)$ is  the trace function, i.e., $g(\bm{P}^i)=\text{trace}(\bm{P}^i)$.  
The algorithm terminates  when the decrease of $g(\bm{P}^{i+1})$ is sufficiently small, i.e.,  $g(\bm{P}^{i})-g(\bm{P}^{i+1})\leq \delta,$ where $\delta$ is a small positive scalar. 
In general, the value of $\delta$ should be chosen on a  case-by-case basis based on prior information or  numerical simulations. 
\begin{remark}
	 Since the sources are static, the proposed method can be extended to multiple measurements in a  straightforward manner using a recursive approach. That is, based on the past measurements, we can derive a bounding set of the state which may be used as the initial value of the algorithm. Moreover, the state bounding set is updated based on the  initial state bounding set and the new measurement. 
\end{remark}
\begin{remark}
	The number of sources has to be known in advance in this paper. In  most studies of the multiple-source localization problem, the number of sources is assumed  known (\cite{Cheng-Wu-Zhang-Wu-Li-Maple12}, \cite{Sheng-Hu05},  \cite{Wang-Quost-Chazot-Antoni16}). When  the number of sources is unknown, the basic idea  is to select an optimization criterion to determine the number of sources.
\end{remark}

\section{Bounding the Remainder}\label{sec_3}
In this section, we consider the problem of determining a bounding box to cover the higher-order remainder.
The bounding box of the remainder is derived at each iteration based on the boundary of the convex bounding set of the state. In particular, when the energy parameter  and the position of each source are bounded in an interval  and a ball respectively, the remainder bound is obtained  analytically.

As shown in Equation (\ref{mfq_1}), the measurement function $f$ is rewritten as a state separable equation:
\begin{align}\label{eqsm_4}
f=\sum_{n=1}^Nf_{n}(\bm{x}_n),
\end{align}
where $f_{n}(\bm{x}_n)=[f_{n,1}(\bm{x}_n),\cdots,f_{n,L}(\bm{x}_n)]^T$, $\bm{x}_n\in \mathbb{R}^{d+1}$ is the state parameter of the $n$-th source. The derivative function of $f$ satisfies $\frac{\partial f(\bm{x})}{\partial \bm{x}_n}=\frac{\partial f_n(\bm{x}_n)}{\partial \bm{x}_n}$. Thus, we have \begin{align}\label{mfq_2}
\begin{split}
\bm{J}(\hat{\bm{x}}^i)&=[\frac{\partial f(\bm{x})}{\partial \bm{x}_1},\cdots,\frac{\partial f(\bm{x})}{\partial \bm{x}_N}]|_{\hat{\bm{x}}^i}\\
&=[\frac{\partial f_1(\bm{x}_1)}{\partial \bm{x}_1}|_{\hat{\bm{x}}_1^i},\cdots,\frac{\partial f_N(\bm{x}_N)}{\partial \bm{x}_N}|_{\hat{\bm{x}}_N^i}].
\end{split}
\end{align}            

The remainder in (\ref{eqsm_11}) is rewritten as
\begin{align}\label{eqsm_13}
\Delta f(\bm{x},\hat{\bm{x}}^i)=f(\bm{x})-f(\hat{\bm{x}})-\bm{J}(\hat{\bm{x}}^i)(\bm{x}-\hat{\bm{x}}^i).
\end{align}
Substituting (\ref{eqsm_4}) and (\ref{mfq_2}) into (\ref{eqsm_13}), the remainder is 
\begin{align}
\Delta f(\bm{x},\hat{\bm{x}}^i)=\sum_{n=1}^N(f_n(\bm{x}_n)-f_n(\hat{\bm{x}}_n^i)-\bm{J}_n(\hat{\bm{x}}_n^i)(\bm{x}_n-\hat{\bm{x}}_n^i)).
\end{align}
where $\bm{J}_n(\hat{\bm{x}}_n^i)=\frac{\partial f_n(\bm{x}_n)}{\partial \bm{x}_n}|_{\hat{\bm{x}}_n^i}$ and $\bm{x}_n\in\mathcal{E}_n^i$, $n=1,\cdots,N$. Denote 
\begin{align}
\begin{split}\label{mfq_4}
\Delta f_{n}(\bm{x}_n,\hat{\bm{x}}_n^i)&=f_n(\bm{x}_n)-\bm{J}_n(\hat{\bm{x}}_n^i)(\bm{x}_n-\hat{\bm{x}}_n^i)\\
&~~~-f_n(\hat{\bm{x}}_n^i),n=1,\cdots,N,
\end{split}
\end{align}
where 
$$\Delta f_{n}(\bm{x}_n,\hat{\bm{x}}_n^i)=[\Delta f_{n,1}(\bm{x}_n,\hat{\bm{x}}_n^i),\cdots,\Delta f_{n,L}(\bm{x}_n,\hat{\bm{x}}_n^i)]^T.$$

If there is a box $\mathcal{B}_n^i$ which contains $\Delta f_{n}(\bm{x}_n,\hat{\bm{x}}_n^i)$ for all $\bm{x}_n\in\mathcal{E}_n^i$, i.e.,
\begin{align}\label{mfq_3}
\begin{split}
\Delta f_{n}(\bm{x}_n,\hat{\bm{x}}_n^i)\in\mathcal{B}_n^i=&\{\bm{z}\in \mathbb{R}^L:\bm{D}_{n,l}^i\leq\bm{z}_l\\
&~~\leq \bm{U}_{n,l}^i,\ l=1,2,\cdots,L\},
\end{split}
\end{align}
where $\bm{D}_{n,l}^i$ is the $l$-th component of the lower bound $\bm{D}_{n}^i$ of the box $\mathcal{B}_n^i$, i.e., $\bm{D}_{n}^i=[\bm{D}_{n,1}^i,\cdots,\bm{D}_{n,L}^i]^T$, $\bm{U}_{n,l}^i$ is the $l$-th component of the upper bound $\bm{U}_{n}^i$ of the box $\mathcal{B}_n^i$, i.e., $\bm{U}_{n}^i=[\bm{U}_{n,1}^i,\cdots,\bm{U}_{n,L}^i]^T$,  
then the bounding box $\mathcal{B}^i$ (see (\ref{eqsm_12})) of the remainder $\Delta f(\bm{x},\hat{\bm{x}}^i)$ is derived with the lower bound and upper bound as follows: 
\begin{align}\label{mfq-33}
\bm{D}^{f,i}=\sum_{n=1}^N\bm{D}_n^i,~\bm{U}^{f,i}=\sum_{n=1}^N\bm{U}_n^i.
\end{align}          

\begin{figure}
\begin{center}
\includegraphics[height=4cm]{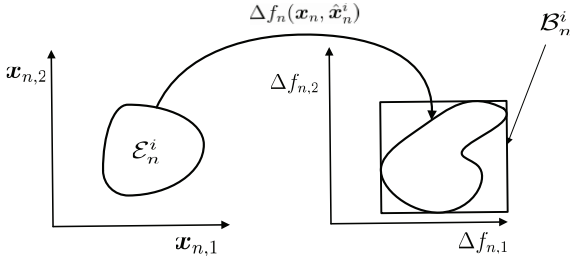}    
\caption{The bounding box $\mathcal{B}_n^i$ of $\Delta f(\bm{x},\hat{\bm{x}}^i)$.}  
\label{Mfig_01}                                 
\end{center}                                 
\end{figure}

The compact bounding box $\mathcal{B}_n^i$ of the remainder set $\{\Delta f_{n}(\bm{x}_n,\hat{\bm{x}}_n^i),\bm{x}_n^i\in\mathcal{E}_n^i\}$, as shown in Fig. \ref{Mfig_01} ($L=2$),  can be equivalently used the following optimization problems, for $l=1,\cdots,L$, 
\begin{align}\label{mfq-16}
\begin{split}
&\max~ t\\
&\text{subject\ to}~ t\leq\Delta f_{n,l}(\bm{x}_n,\hat{\bm{x}}_n^i),\ \bm{x}_n\in\mathcal{E}_n^i,
\end{split}
\end{align}
and 
\begin{align}\label{mfq-17}
\begin{split}
&\min~ t\\
&\text{subject\ to}~ t\geq\Delta f_{n,l}(\bm{x}_n,\hat{\bm{x}}_n^i),\ \bm{x}_n\in\mathcal{E}_n^i. 
\end{split}
\end{align}
Since there are infinite number of constraints, the problem (\ref{mfq-16})-(\ref{mfq-17}) is a  semi-infinite optimization problem \cite{Boyd-Vandenberghe04}. In general, it is an NP-hard problem. In order to  reduce the computational complexity,
we have the following result on finding the bounds of the remainder. 
\begin{prop}\label{proposition_1_1}
	At $i$-th iteration, 
	 the parameter of the $n$-th source is contained in a closed convex set $\mathcal{E}_n^i$, i.e.,    $[s_n,\rho_n^T]^T\in \mathcal{E}_n^i$ defined in (\ref{eqsm_14}),  the bounds of the remainder $\Delta f_{n,l}(\bm{x}_n,\hat{\bm{x}}_n^i)$ are obtained as follows:
	 
	(a) If the $l$-th sensor is not contained in the set $\mathcal{E}_n^{\rho,i}$, then the minimum and maximum of $\Delta f_{n,l}(\bm{x}_n,\hat{\bm{x}}_{n}^i)$ are  obtained at the stationary point $\hat{\bm{x}}_n^i$ or  on the boundary of $\mathcal{E}_n^{i}$.
	
	(b) If the  $l$-th sensor is contained in the set $\mathcal{E}_n^{\rho,i}$, then the maximum of   $\Delta f_{n,l}(\bm{x}_n,\hat{\bm{x}}_{n}^i)$ is $+\infty$ and  the minimum  $\Delta f_{n,l}(\bm{x}_n,\hat{\bm{x}}_{n}^i)$ is  obtained  at the stationary point $\hat{\bm{x}}_n^i$  or  on the boundary of $\mathcal{E}_n^{i}$.
\end{prop}
\begin{pf}
 See the Appendix.
\end{pf}

\begin{remark}
	Proposition \ref{proposition_1_1} means that when we determine the remainder bound,  only the boundary of the set $\mathcal{E}_n^i$ and the stationary point $\hat{\bm{x}}_n^i$ are useful. It is not necessary to consider the other interior points of the set $\mathcal{E}_n^i$ except the stationary point $\hat{\bm{x}}_n^i$. Thus, the computational complexity is reduced  quite significantly.  When we take samples from the boundary of the set $\mathcal{E}_n^i$, they are sufficient to derive the outer bounding box of the remainder set.
\end{remark}
\begin{remark}
	To guarantee that the resulting box actually contains the true remainder set, we can heuristically enlarge the sampling area, such as taking samples from the boundary of the larger set $1.1\cdot\mathcal{E}_n^i=1.1\cdot\mathcal{E}_{n}^{s,i}\times1.1\cdot\mathcal{E}_{n}^{\rho,i}$, where 
	$
	1.1\cdot\mathcal{E}_{n}^{s,i}=\{s_n\in\mathbb{R}^{1}: |s_n-\hat{s}_n^i|\leq 1.1\cdot S_n^i\},$
	$1.1\cdot\mathcal{E}_{n}^{\rho,i}=\{\rho\in\mathbb{R}^{d}: (\rho-\hat{\rho}_{n}^i)^T (\bm{P}_{n}^i)^{-1}(\rho-\hat{\rho}_{n}^i)\leq(1.1)^2\},
	$
    then the remainder set becomes a little larger than that based on $\bm{x}_n\in\mathcal{E}_{n}^i$. If we derive
	a box to cover the little larger remainder, then this box can cover the original remainder set $\{\Delta f_{n}(\bm{x}_n,\hat{\bm{x}}_n^i),\bm{x}_n\in\mathcal{E}_n^i\}$. 
\end{remark}

Furthermore,  if $s_n$ is  contained in the interval $\mathcal{E}_{n}^{s,i}=\{s_n\in\mathbb{R}^{1}: |s_n-\hat{s}_n^i|\leq S_n^i\}$ and $\rho_n$ is  contained in the ball $\mathcal{E}_{n}^{\rho,i}=\{\rho_n\in\mathbb{R}^{d}: \|\rho_n-\hat{\rho}_n^i\|\leq R_n^i\}$ as defined in (\ref{eqsm_14}), then  the remainder bound is obtained  analytically  as stated in the  following propositions.
\begin{prop}\label{proposition_1}
	If the energy parameter and the position of the source are bounded
	in an interval and  a ball respectively, i.e., $\bm{x}_n\in \mathcal{E}_n^i=\mathcal{E}_{n}^{s,i}\times\mathcal{E}_{n}^{\rho,i}$, $\mathcal{E}_{n}^{s,i}=\{s_n\in\mathbb{R}^{1}: |s_n-\hat{s}_n^i|\leq S_n^i\}$ and $\mathcal{E}_{n}^{\rho,i}=\{\rho_n\in\mathbb{R}^{d}: \|\rho_n-\hat{\rho}_n^i\|\leq R_n^i\}$, and  the sensors are not contained in the state bounding set, i.e., $R_n^i<\min\{\tau_l=\|\hat{\rho}_n^i-r_l\|:\ l=1,\cdots,L\}$, where $r_l$, $\hat{\rho}_n^i$  and $\alpha>0$ are defined in (\ref{eqsm_2}), then the bounding box $\mathcal{B}_n^i$ of the remainder $\Delta f_{n}(\bm{x}_n,\hat{\bm{x}}_n^i)$, is obtained  analytically, i.e., the maximum and minimum of function $\Delta f_{n,l}(\bm{x}_n,\hat{\bm{x}}_n^i)$, for $l=1,\cdots,L$, are
	\begin{align}
	\begin{split}
	\max_{\bm{x}_n\in \mathcal{E}_n^i}f_{n,l}(\bm{x}_n,\hat{\bm{x}}_n^i)&=\max\bigg\{\Delta  \tilde{f}_{n,l}(1,-S_n^i,R_n^i),\\
	\Delta  \tilde{f}_{n,l}(-1,S_n^i,R_n^i)&,\Delta  \tilde{f}_{n,l}(-1,-S_n^i,R_n^i),0\bigg\},
	\end{split}
	\end{align}
	\begin{align}
	\begin{split}
	\min_{\bm{x}_n\in \mathcal{E}_n^i}\Delta f_{n,l}(\bm{x}_n,\hat{\bm{x}}_n^i)&=\min\bigg\{\Delta  \tilde{f}_{n,l}(-1,-S_n^i,t_1), \\
	\Delta  \tilde{f}_{n,l}(1,S_n^i,t_2), \Delta  &\tilde{f}_{n,l}(\max\{-1,\hat{k}_1\},-S_n^i,R_n^i),\\
	 \Delta  \tilde{f}_{n,l}(\max\{-1,&\min\{\hat{k}_2,1\}\},S_n^i,R_n^i),0\bigg\},
	\end{split}
	\end{align}
	where
	\begin{align*}
	\Delta \tilde{f}_{n,l}(k,\Delta s_n,t)
	&=g_l\bigg(\frac{\hat{s}_n^i+\Delta s_n}{(t^2+\tau_l^2+2t\tau_l k)^{\alpha/2}}+\frac{\alpha\hat{s}_n^itk}{\tau_l^{\alpha+1}}\\
	&-\frac{\hat{s}_n^i+\Delta s_n}{\tau_l^\alpha}\bigg),
	\end{align*}
	is a function of variables $k,\Delta s$ and $t$.
	\begin{align}
	t_1&=\min\{\tau_l\bigg(1-(1-\frac{S_n^i}{\hat{s}_n^i})^{1/(\alpha+1)}\bigg),R_n^i\},\label{mfq-23}\\
	t_2&=\min\{\tau_l\bigg((1+\frac{S_n^i}{\hat{s}_n^i})^{1/(\alpha+1)}-1\bigg),R_n^i\},\label{mfq-24}\\
	\hat{k}_1&=\frac{\bigg((1-\frac{S_n^i}{\hat{s}_n^i})^{1/(\alpha/2+1)}-1\bigg)\tau_l^2-(R_n^i)^2}{2R_n^i\tau_l},\label{mfq-25}\\
	\hat{k}_2&=\frac{\bigg((1+\frac{S_n^i}{\hat{s}_n^i})^{1/(\alpha/2+1)}-1\bigg)\tau_l^2-(R_n^i)^2}{2R_n^i\tau_l}\label{mfq-26}.
	\end{align}    
\end{prop}
\begin{pf}
	See the Appendix.
\end{pf}
\begin{remark}
	Proposition \ref{proposition_1} means that when the energy parameter and the position of the source
	are bounded in an interval and a ball  respectively,  the bounding box
	of the remainder  is obtained  analytically. The upper and lower bounds of the bounding box $\mathcal{B}^i_n$, for $l=1,\cdots,L$, are
	\begin{align*}
	\bm{D}_{n,l}^i&=\min_{\bm{x}_n\in \mathcal{E}_n}\Delta f_{n,l}(\bm{x}_n,\hat{\bm{x}}_n^i),\\
	\bm{U}_{n,l}^i&=\max_{\bm{x}_n\in \mathcal{E}_n}\Delta f_{n,l}(\bm{x}_n,\hat{\bm{x}}_n^i).
	\end{align*}
	Obviously, the computational complexity of finding the remainder bound is significantly reduced due to the availability of the analytical solution.
\end{remark}

In Proposition \ref{proposition_1}, we have assumed that the set $\mathcal{E}_{n}^{\rho,i}$ does not contain any sensor.  If this assumption is not satisfied, we have the following result.  
\begin{prop}\label{proposition_1_c}
	If the energy parameter and the position of the source are bounded
	in an interval and  a ball respectively, as shown in Proposition \ref{proposition_1}, and  the $l$-th sensor is contained in the state bounding set, i.e., $R_n>\tau_l=\|\hat{\rho}_n^i-r_l\|$, where $r_l$, $\hat{\rho}_n^i$  and $\alpha>0$ are defined in (\ref{eqsm_2}),  the  minimum of the remainder $\Delta f_{n,l}(\bm{x}_n,\hat{\bm{x}}_n^i)$ is
	\begin{align}
	\begin{split}
	\min_{\bm{x}_n\in \mathcal{E}_n^i}&\Delta f_{n,l}(\bm{x}_n,\hat{\bm{x}}_n^i)=g_l\min\bigg\{\Delta \tilde{f}_{n,l}(-1,-S_n^i,t_1^-), \\
	&~~~~~~~~~~\Delta \tilde{f}_{n,l}(1,S_n^i,t_2^-),\Delta \tilde{f}_{n,l}(\hat{k}_1,-S_n^i,R_n^i),\\
	&~~~~~~~~~~\Delta \tilde{f}_{n,l}(\hat{k}_2,S_n^i,R_n^i),0\bigg\},
	\end{split}
	\end{align}
	where
	$\Delta \tilde{f}_{n,l}(k,\Delta s,t)$, $\hat{k}_1$ and  $\hat{k}_2$ are same as that in Proposition \ref{proposition_1}, and
	\begin{align*}
	t_1^-&=\tau_l\bigg(1-(1-\frac{S_n^i}{\hat{s}_n^i})^{1/(\alpha+1)}\bigg),\\
	t_2^-&=\tau_l\bigg((1+\frac{S_n^i}{\hat{s}_n^i})^{1/(\alpha+1)}-1\bigg).
	\end{align*}         
\end{prop}
\begin{pf}
See the Appendix.
\end{pf}
\begin{remark}
	It is easy to find that when $R_n>\tau_l$, $\max_{\bm{x}_n\in \mathcal{E}_n^i}\Delta f_{n,l}(\bm{x}_n,\hat{\bm{x}}_n^i)=+\infty$. It means that the remainder cannot be covered by a bounded box. In this case, the remainder is constrained by a hyperplane. 
\end{remark}

\section{Multiple-source Ellipsoidal Localization Algorithm}\label{sec_4}
In this section, we derive the multiple-source ellipsoidal localization  method.
The main idea  is that based on the separability of the nonlinear measurement function, an S-procedure estimation method is developed to deal with the multiple-source localization problem  by using an alternating optimization iterative algorithm.

For multiple-source localization, the bounding box $\mathcal{B}_{n}^i$ (see (\ref{mfq_3})) of the remainder $\Delta f_{n}$, for $n=1,\ldots,N$, 
is derived based on the current bounding set of the $n$-th source state by Proposition \ref{proposition_1_1} or Propositions \ref{proposition_1}-\ref{proposition_1_c}, at the $i$-th iteration.
The bounding box $\mathcal{B}^{i}$ of the remainder $\Delta f$ 
is derived based on the bounding boxes $\mathcal{B}_{n}^i$, $n=1,\cdots,N$, as shown in (\ref{mfq-33}).  The set $\{1,\cdots,L\}$ is divided into two disjoint subsets $\mathcal{L}^{+,i}$ and $\mathcal{L}^{-,i}$,
\begin{align}
\mathcal{L}^{+,i}=\{l:\bm{U}^{f,i}_l<+\infty,l=1,\cdots,L\},\\
\mathcal{L}^{-,i}=\{l:\bm{U}^{f,i}_l=+\infty,l=1,\cdots,L\}.
\end{align}

Moreover, we use the current state bound $\mathcal{E}_{1}^i\times\cdots\times\mathcal{E}_{N}^i$ and the remainder bound $\mathcal{B}_{1}^i\times$ $\cdots\times\mathcal{B}_{N}^i$ to determine the bounding set of the state at $(i+1)$-th iteration, i.e., look for $\hat{\rho}_{n}^{i+1}$, $\hat{s}_{n}^{i+1}$,  $S_{n}^{i+1}$ and $\bm{P}_{n}^{i+1}$ of $\mathcal{E}_{n}^{i+1}$ such that the state $\bm{x}_{n}$ of the $n$-th source  belongs to $\mathcal{E}_{n}^{i+1}$, $n=1,\cdots,N$.
It is obtained by the following proposition.

\begin{prop}\label{proposition_2}
	At ($i+1$)-th  iteration, based on measurement $\bm{y}$, the current state bound $\mathcal{E}_{1}^i\times\cdots\times\mathcal{E}_{N}^i$, the current remainder bound $\mathcal{B}_{1}^i\times\cdots\times\mathcal{B}_{N}^i$,  and the noise bounding box $\mathcal{B}^{\varepsilon}$,   for the $n$-th source state $\bm{x}_{n}$, $n= 1,\cdots, N$, we have:
	
	(a) The state bounding set $\mathcal{E}_{n}^{s,(i+1)}$, as shown in (\ref{eqsm_9}), is obtained by solving the optimization problem in the variables $(S_{n}^{i+1})^{2},\ \hat{s}_{n}^{i+1}$, nonnegative scalars $\tau_{j}^{1}\geq0$,  $\tau_{j}^{2}\geq0$, $j= 1,\cdots, N,$  $\tau_{l}^{+}\geq0$, $l\in\mathcal{L}^{+,i}$, $\tau_{l}^{-}\geq0$, $l\in\mathcal{L}^{-,i}$,
	\begin{align}
	\label{therom_eqsm_33}&\min\ \  (S_{n}^{i+1})^2\\
	&\nonumber\text{subject\ to}\ \\
	\begin{split}\label{therom_eqsm_34}
	&-\tau_{j}^{1}\leq0, -\tau_{j}^{2}\leq0, j= 1,\cdots, N,\\  &-\tau_{l}^{+}\leq0,l\in\mathcal{L}^{+,i},
	-\tau_{l}^{-}\leq0,\ l\in\mathcal{L}^{-,i},
	\end{split}
	\end{align}
	\begin{multline}\label{therom_eqsm_35}
	\left[\begin{array}{c}
	-(S_{n}^{i+1})^2\\
	(\Phi_{n}^{s,(i+1)}(\Psi^{+,(i+1)})_{\bot})^T\\
	\end{array}\right.\\
	\left.\begin{array}{c}
	\Phi_{n}^{s,(i+1)}(\Psi^{+,(i+1)})_{\bot}\\
	-(\Psi^{+,(i+1)})_{\bot}^T\Xi_{n}(\Psi^{+,(i+1)})_{\bot}\\
	\end{array}\right]\preceq0, 
	\end{multline}
	where
	\begin{align}
	\label{mfq-34}\Phi_{n}^{s,(i+1)}&=[\hat{s}_{n}^i-\hat{s}_{n}^{i+1},\bm{I}_{n}^1\hat{\bm{E}}^{i},0,0],\\
\label{mfq-35}
\begin{split}
\Psi^{+,(i+1)}&=[f^+(\hat{\bm{x}}^{i})+\hat{\bm{e}}^{+,i}\bm{e}^{\varepsilon,+}-\bm{y}^+,\bm{J}^{+,i}\hat{\bm{E}}^{i},\\
&\ \ \ \ \ \text{diag}(\frac{\hat{\bm{b}}^{+,i}+\bm{b}^{\varepsilon,+}}{2})],
\end{split}\\	
\label{mfq-36}	\Psi_{l}^{-,(i+1)}&=\begin{bmatrix}
	H_l&\frac{1}{2}\bm{J}_{l}^{i}\hat{\bm{E}}^{i}&\bm{0}\\
	\frac{1}{2}(\bm{J}_{l}^{i}\hat{\bm{E}}^{i})^T&\bm{0}&\bm{0}\\
	\bm{0}&\bm{0}&\bm{0}
	\end{bmatrix},\\
	\begin{split}
	H_l&=f_{l}(\hat{\bm{x}}^{i})+\bm{e}^{\varepsilon}_l-\frac{b^{\varepsilon}_l}{2}+D^{f,i}_l-\bm{y}_{l},~~ l\in\mathcal{L}^{-,i},
	\end{split}
	\end{align}
	$\bm{I}_{n}^1=[\overbrace{0,\cdots,\bm{I}_1,\cdots,0}^{the\ n-th\ entry\ is\ \bm{I}_1}]$, $\bm{I}_1=\begin{bmatrix}
	1&\mathbf{0}\\
	\end{bmatrix}$,   $\mathbf{0}\in \mathbb{R}^{1\times d}$, $\hat{\bm{E}}^{i}=\text{diag}\{s_{1}^i,\bm{E}_{1}^i,\ldots,s_{N}^i,\bm{E}_{N}^i\}$ is the block diagonal matrix of $s_{j}^i$ and Cholesky factorization $\bm{E}_{j}^i$,  $\bm{P}_{j}^i=\bm{E}_{j}^i(\bm{E}_{j}^i)^T$,  $\hat{\bm{e}}^{+,i}_m=\frac{\bm{D}^{f,i}_{l_m}+\bm{U}^{f,i}_{l_m}}{2}$, $\hat{\bm{b}}^{+,i}_m=\frac{\bm{U}^{f,i}_{l_m}-\bm{D}^{f,i}_{l_m}}{2}$, $l_m\in\mathcal{L}^{+,i}$, as shown in (\ref{mfq-33}),  $\bm{e}^{\varepsilon}=\frac{D^\varepsilon+U^\varepsilon}{2},\ \bm{b}^{\varepsilon}=\frac{U^\varepsilon-D^\varepsilon}{2}$ are shown in (\ref{eqsm_5}), $(\Psi^{+,(i+1)})_\bot$ is the orthogonal complement of $\Psi^{+,(i+1)}$ with full column rank, i.e., a basis of the null space of $\Psi^{+,(i+1)}$, $f^+=[f_{l_1},\cdots,f_{l_{L^+}}]^T,~l_m\in\mathcal{L}^{+,i},~m=1,\cdots,L^+$,  $\bm{J}^{+,i}=\frac{\partial f^+(x)}{\partial x}|_{\hat{\bm{x}}^{i}}$,  $L^+=|\mathcal{L}^{+,i}|$.  The symbol $\preceq$ is used  to denote generalized inequality between symmetric matrices, it represents
	matrix inequality.
	\begin{align}
	\begin{split}\label{mfq-37}
     \Xi_{n}&=\text{diag}\bigg(1-\sum_{j=1}^N\tau_{j}^{1}-\sum_{j=1}^N\tau_{j}^{2}-\sum_{l\in\mathcal{L}^{+,i}}\tau_{l}^{+},\\
     &~~~~~~~\text{diag}(\tau_{1}^{1},\tau_{1}^{2}\bm{I},\cdots,\tau_{N}^{1},\tau_{N}^{2}\bm{I}),\\
     &~~~~~~~\text{diag}(\tau_{l_1}^{+},\cdots,\tau_{l_{L^+}}^{+})\bigg)+\sum_{l\in\mathcal{L}^{-,i}}\tau_{l}^{-}\Psi_{l}^{-,(i+1)}.
	\end{split}
	\end{align}
	
	(b) The state bounding set $\mathcal{E}_{n}^{\rho,(i+1)}$, as shown in (\ref{eqsm_9}), is obtained by solving the optimization problem in the variables $\bm{P}_{n}^{i+1},\ \hat{\rho}_{n}^{i+1}$, nonnegative scalars $\tau_{j}^{1}\geq0$,  $\tau_{j}^{2}\geq0$, $j= 1,\cdots, N,$   $\tau_{l}^{+}\geq0$, $l\in\mathcal{L}^{+,i}$, $\tau_{l}^{-}\geq0$, $l\in\mathcal{L}^{-,i}$,
	\begin{align}
	\label{therom_eqsm_33_2}&\min\ \  g(\bm{P}_{n,i+1})\\
	&\nonumber\text{subject\ to}\ \\
	\begin{split}\label{therom_eqsm_34_2}
	&-\tau_{j}^{1}\leq0, -\tau_{j}^{2}\leq0, j= 1,\cdots, N,\\  &-\tau_{l}^{+}\leq0,l\in\mathcal{L}^{+,i},
	-\tau_{l}^{-}\leq0,\ l\in\mathcal{L}^{-,i},
	\end{split}
	\end{align}
	\begin{multline}\label{therom_eqsm_35_2}
	\left[\begin{array}{c}
	-\bm{P}_{n}^{i+1}\\
	(\Phi^{\rho,(i+1)}_n(\Psi^{+,(i+1)})_{\bot}^T\\
	\end{array}\right.\\
	\left.\begin{array}{c}
	\Phi^{i+1}_2(\Psi^{+,(i+1)})_{\bot}\\
	-(\Psi^{+,(i+1)})_{\bot}^T\Xi_{n}(\Psi^{+,(i+1)})_{\bot}\\
	\end{array}\right]\preceq0, 
	\end{multline}
	where
	\begin{align}
	\Phi^{\rho,(i+1)}_n&=[\hat{\rho}_{n}^i-\hat{\rho}_{n}^{i+1},\bm{I}_{n}^2\hat{\bm{E}}^{i},0,0],
	\end{align}
	$\bm{I}_{n}^2=[\overbrace{0,\cdots,\bm{I}_2,\cdots,0}^{the\ n-th\ entry\ is\ \bm{I}_2}]$, $\bm{I}_2=\begin{bmatrix}
	\mathbf{0}&\bm{I}
	\end{bmatrix}$, $\mathbf{0}\in \mathbb{R}^{d\times 1}$, $\bm{I}\in \mathbb{R}^{d\times d}$ and other symbols are same as in problem (\ref{therom_eqsm_33})-(\ref{therom_eqsm_35}).
\end{prop}
\begin{pf}
	See the Appendix.
\end{pf}
\begin{remark}
	In the problem (\ref{therom_eqsm_33})-(\ref{therom_eqsm_35}), $\mathcal{E}_{n}^{s,(i+1)}$ is estimated while $\mathcal{E}_{n}^{\rho}$ and the parameter bounds of the other sources are fixed.
	In the problem (\ref{therom_eqsm_33_2})-(\ref{therom_eqsm_35_2}), $\mathcal{E}_{n}^{\rho,(i+1)}$ is estimated while $\mathcal{E}_{n}^s$ and the parameter bounds of the other sources remain  fixed.
	All the problems are feasible. Moreover, $\mathcal{E}_{n}^{s,i}$ and $\mathcal{E}_{n}^{\rho,i}$ are the feasible solutions of the two problems, respectively.
	The problem (\ref{therom_eqsm_33})-(\ref{therom_eqsm_35}) is a convex SDP problem involving a constraint matrix of dimension $M_1=(d+1)N+L+2$, and $M_2=2+2N+2L$ decision variables. Therefore, using a general purpose primal-dual interior-point SDP solver, the practical complexity is $O(M_1^2M_2^2)$. In our context, this corresponds to  $O(d^2)$ where $d$ is the dimension  of the state $\rho$, $O(N^4)$ where $N$ is  the number of sources, and $O(L^4)$ where $L$ is  the number of sensors. In the same way, the  complexity of the problem (\ref{therom_eqsm_33_2})-(\ref{therom_eqsm_35_2}) corresponds to  $O(d^6)$ where $d$ is the dimension  of $\rho$, $O(N^4)$  where $N$ is  the number of sources, and $O(L^4)$  where $L$ is  the number of sensors. Moreover, the decoupled technique in \cite{Calafiore-ElGhaoui04}  can reduce the complexity when the dimension of the state is greater than one. 
\end{remark}

\begin{algorithm}[htb]
	\caption{Ellipsoidal localization  algorithm}
	\label{alg_1}
	\begin{algorithmic}[1]
		\Require
		$g(\bm{P})$: objective function,
		$\mathcal{E}_{1}^0\times$$\cdots\times\mathcal{E}_{N}^0$: initial state bounding set,
		$\mathcal{B}^{\varepsilon}$: the bounding set of noises,
		$\mathcal{B}_{1}^0\times\cdots\times\mathcal{B}_{N}^0$: the bounding set of remainder,
		$\delta>0$: tolerance.
		\Repeat
		\State{\textbf{Input}: the current state bound $\prod_{n=1}^N\mathcal{E}_{n}^i$, and the current remainder bound $\prod_{n=1}^N\mathcal{B}_{n}^i$.}
		\For{$n=1,\cdots,N$}
		\begin{enumerate}
			\item Optimize the center and shape matrix of the state bounding ellipsoid $\mathcal{E}_{n}^{\rho,(i+1)}$ by the problem (\ref{therom_eqsm_33_2})-(\ref{therom_eqsm_35_2}) in Proposition \ref{proposition_2}.
			\item Optimize the center and shape matrix of the state bounding ellipsoid $\mathcal{E}_{n}^{s,(i+1)}$ by the problem (\ref{therom_eqsm_33})-(\ref{therom_eqsm_35}) in Proposition \ref{proposition_2}.
			\item Derive the bounding set $\mathcal{B}_{n}^{i+1}$, by Proposition \ref{proposition_1_1}.
			\item \textbf{Update}: the current state bound $\mathcal{E}_{1}^{i+1}\times\cdots\times\mathcal{E}_{n}^{i+1}\times \mathcal{E}_{n+1}^i\times\cdots\times\mathcal{E}_{N}^i$ and the current remainder bound $\mathcal{B}_{1}^{i+1}\times$ $\cdots$ $\times\mathcal{B}_{n}^{i+1}\times$ $\mathcal{B}_{n+1}^i\times$ $\cdots$ $\times\mathcal{B}_{N}^i$.
		\end{enumerate}
		\EndFor
		\State \textbf{Output}: the current state bound $\mathcal{E}_{1}^{i+1}\times\cdots\times\mathcal{E}_{N}^{i+1}$, and the current remainder bound $\mathcal{B}_{1}^{i+1}\times$ $\cdots$ $\times\mathcal{B}_{N}^{i+1}$.
		\Until{($g(\bm{P}^{i})-g(\bm{P}^{i+1})\leq \delta$)}
	\end{algorithmic}
\end{algorithm}
Using Propositions \ref{proposition_1_1} and   \ref{proposition_2} , we have the alternating optimization iterative algorithm, Algorithm \ref{alg_1}, for the multiple-source localization problem. Moreover, the computational complexity of finding the remainder bound in Propositions \ref{proposition_1} and \ref{proposition_1_c} is  greatly reduced compared to that of   Proposition \ref{proposition_1_1}.
In order to reduce the complexity, the ellipsoidal state  bound $\mathcal{E}^{\rho}$ is relaxed to a bounding ball where the radius is the long semi-axis of the ellipsoid $\mathcal{E}^{\rho}$.
Thus, using Propositions \ref{proposition_1}-\ref{proposition_2}, we can get the Algorithm \ref{alg_2}.
\begin{algorithm}[htb]
	\caption{Simplified ellipsoidal localization  algorithm}
	\label{alg_2}
	\begin{algorithmic}[1]
		\Require
		$g(\bm{P})$: objective function,
		$\mathcal{E}_{1}^0\times$$\cdots\times\mathcal{E}_{N}^0$: initial state bounding set,
		$\mathcal{B}^{\varepsilon}$: the bounding set of noises,
		$\mathcal{B}_{1}^0\times\cdots\times\mathcal{B}_{N}^0$: the bounding set of remainder,
		$\delta>0$: tolerance.
		\Repeat
		\State{\textbf{Input}: the current state bound $\prod_{n=1}^N\mathcal{E}_{n}^i$, and the current remainder bound $\prod_{n=1}^N\mathcal{B}_{n}^i$.}
		\For{$n=1,\cdots,N$}
		\begin{enumerate}
			\item Optimize the center and shape matrix of the state bounding ellipsoid $\mathcal{E}_{n}^{\rho,(i+1)}$ by the problem (\ref{therom_eqsm_33_2})-(\ref{therom_eqsm_35_2}) in Proposition \ref{proposition_2}.
			\item Optimize the center and shape matrix of the state bounding ellipsoid $\mathcal{E}_{n}^{s,(i+1)}$ by the problem (\ref{therom_eqsm_33})-(\ref{therom_eqsm_35}) in Proposition \ref{proposition_2}.
			\item Find the minimum ball which contains  $\mathcal{E}_{n}^{\rho,(i+1)}$. Derive the bounding set $\mathcal{B}_{n}^{i+1}$, by Propositions \ref{proposition_1} and \ref{proposition_1_c}.
			\item \textbf{Update}: the current state bound $\mathcal{E}_{1}^{i+1}\times\cdots\times\mathcal{E}_{n}^{i+1}\times \mathcal{E}_{n+1}^i\times\cdots\times\mathcal{E}_{N}^i$ and the current remainder bound $\mathcal{B}_{1}^{i+1}\times$ $\cdots$ $\times\mathcal{B}_{n}^{i+1}\times$ $\mathcal{B}_{n+1}^i\times$ $\cdots$ $\times\mathcal{B}_{N}^i$.
		\end{enumerate}
		\EndFor
		\State \textbf{Output}: the current state bound $\mathcal{E}_{1}^{i+1}\times\cdots\times\mathcal{E}_{N}^{i+1}$.
		\Until{($g(\bm{P}^{i})-g(\bm{P}^{i+1})\leq \delta$)}
	\end{algorithmic}
\end{algorithm}

\begin{remark}
	Algorithm \ref{alg_1} and Algorithm \ref{alg_2} are similar to the block coordinate decent or nonlinear Gauss-Seidel methods. At each iteration, the objecive  function $g(\bm{P}^{i+1})$ is minimized with respect to each of the ``block coordinate'' vectors $\bm{x}_{n}$,  in a cyclic manner. The criterion for terminating the iterations is that the algorithm stops  when the decrease of $g(\bm{P}^{i+1})$ is sufficiently small, i.e., $g(\bm{P}^{i})-g(\bm{P}^{i+1})\leq \delta,$ where $\delta$ is a small positive scalar. This method can  converge to a stationary point. A detailed discussion of the method is found in \cite{Bertsekas99}, \cite{Zhu03}.
\end{remark}
\begin{remark}
	 When measurement data may contain outliers, the guaranteed outlier minimal number estimator  (GOMNE) can be used to deal with the problem  \cite{Jaulin-Kieffer-Walter-Meizel02}. Moreover,  gating is a screening  technique that proves very effective in cutting down the number of unlikely tracks postulated for a target (\cite{Bar-Shalom90}, \cite{BarShalom-Li-Kirubarajan01}). The idea of gating can also be used  to delete outliers.
\end{remark}
\section{$\alpha$ is unknown but bounded }\label{sect-1}

In this section, we derive the multiple-source ellipsoidal localization  method when $\alpha$ is unknown but bounded. The main idea  is that based on  the separability of the nonlinear measurement function and interval mathematics, an efficient estimation procedure  is developed to deal with the multiple-source localization problem by using an alternating
optimization iterative algorithm.

Consider the multiple source localization problem when  $\alpha$ is unknown but bounded. Since the decay factor $\alpha$ usually lies between 2 to 4 (see \cite{Meng-Xiao-Xie11}), we assume  that $\alpha$ is bounded and lies in  $[2,4]$, i.e., $\alpha\in[\alpha_{1},\alpha_{2}]\subset[2,4]$.   
The measurement functions are written as  
\begin{align}
y_l=\sum_{n=1}^Nf_{n,l}(\bm{x}_n)+\varepsilon_l,~l=1,\cdots,L, 
\end{align}
where $f_{n,l}(\bm{x}_n)$ is defined in (\ref{mfq_1}). 
At the $i$-th iteration, the bounding set of the state $\bm{x}$ is $\mathcal{E}^i$, which is defined in (\ref{eqsm_9}). The bounding interval $[\breve{D}_{n,l}^i,\breve{U}_{n,l}^i]$ of the function $f_{n,l}$ can be obtained. 
\begin{lem}\label{lemma-1}
If the energy parameter and the position of the $n$-th source are bounded in  $\mathcal{E}_n^i$ which is defined in (\ref{eqsm_9}),  and $\alpha\in[\alpha_{1},\alpha_{2}]\subset[2,4]$, then the bounding interval $[\breve{D}_{n,l}^i,\breve{U}_{n,l}^i]$ of the function $f_{n,l}$, is obtained by solving the optimization problems
\begin{align}\label{2019-6-8}
\begin{split}
&D_{n,l}^{\rho}=\min \|\rho_n-r_l\|\\
&\text{subject\ to\ } \rho_{n}\in\mathcal{E}_{n}^{\rho,i},
\end{split}
\end{align}
and
\begin{align}\label{2019-5-9-4}
\begin{split}
&U_{n,l}^{\rho}=\max \|\rho_n-r_l\|\\
&\text{subject\ to\ } \rho_{n}\in\mathcal{E}_{n}^{\rho,i}.
\end{split}
\end{align}
If $D_{n,l}^{\rho}>0$,
\begin{align*}
[\breve{D}_{n,l}^i,\breve{U}_{n,l}^i]=&[g_l\frac{\hat{s}_{n}^i-S_{n}^i}{\max\{(U_{n,l}^{\rho})^{\alpha_{1}},(U_{n,l}^{\rho})^{\alpha_{2}}\}},\\
&~~~~~~~~~g_l\frac{\hat{s}_{n}^i+S_{n}^i}{\min\{(D_{n,l}^{\rho})^{\alpha_{1}},(D_{n,l}^{\rho})^{\alpha_{2}}\}}].
\end{align*}
If $D_{n,l}^{\rho}=0$,
\begin{align*}
[\breve{D}_{n,l}^i,\breve{U}_{n,l}^i]=[g_l\frac{\hat{s}_{n}^i-S_{n}^i}{\max\{(U_{n,l}^{\rho})^{\alpha_{1}},(U_{n,l}^{\rho})^{\alpha_{2}}\}},+\infty], 
\end{align*}
where $\hat{s}_{n}^i$ and $S_{n}^i$ are defined in (\ref{eqsm_9}).
\end{lem}
\begin{pf}
	See the Appendix.
\end{pf}
\begin{remark}
	The optimization problem (\ref{2019-6-8}) is a second-order cone program (SOCP) and it can be solved by using Yalmip \cite{Lofberg04}. Using the S-procedure method, the optimization problem
	(\ref{2019-5-9-4}) is  equivalent to the  following optimization problem 
	\begin{align}
	\begin{split}
	&\min t\\
	&\text{subject\ to\ } \tau\geq0\\
	&\begin{bmatrix}
	\bm{I}&-r_l\\
	-r_l^T&r_l^Tr_l-t
	\end{bmatrix}\\
	&~~~~~~~-\tau\begin{bmatrix}
	(\bm{P}_{n}^i)^{-1}&-(\bm{P}_{n}^i)^{-1}\hat{\rho}_{n}^i\\
	-((\bm{P}_{n}^i)^{-1}\hat{\rho}_{n}^i)^T&(\hat{\rho}_{n}^i)^T(\bm{P}_{n}^i)^{-1}\hat{\rho}_{n}^i-1
	\end{bmatrix}\preccurlyeq0, 
	\end{split}
	\end{align}
	where $\bm{P}_{n}^i$ and $\hat{\rho}_{n}^i$ are defined in (\ref{eqsm_9}).  
\end{remark}

We use the current state bound $\mathcal{E}_{1}^i\times\cdots\times\mathcal{E}_{N}^i$ and the bounding intervals $[\breve{D}_{n,l}^i,\breve{U}_{n,l}^i]$ of the function $f_{n,l}$, $l=1,\cdots,L,n=1,\cdots,N$, to determine the bounding set of the state at $(i+1)$-th iteration, i.e., look for $\hat{\rho}_{n}^{i+1}$, $\hat{s}_{n}^{i+1}$,  $S_{n}^{i+1}$ and $\bm{P}_{n}^{i+1}$ of $\mathcal{E}_{n}^{i+1}$ such that the state $\bm{x}_{n}$ of the $n$-th source  belongs to $\mathcal{E}_{n}^{i+1}$, $n=1,\cdots,N$.
It is obtained by the following proposition.

\begin{prop}\label{proposition_3}
At ($i+1$)-th  iteration, based on measurement $\bm{y}$, the current state bound $\mathcal{E}_{1}^i\times\cdots\times\mathcal{E}_{N}^i$, $\alpha\in[\alpha_{1},\alpha_{2}]\subset[2,4]$,  the bounding  interval $[\breve{D}_{n,l}^i,\breve{U}_{n,l}^i]$ of the  function $f_{n,l}$,  $l=1,\cdots,L,n=1,\cdots,N$, and the noise bounding box $\mathcal{B}^{\varepsilon}$, for the $n$-th source state $\bm{x}_{n}$, $n= 1,\cdots, N$, we have:

(a) The state bounding set $\mathcal{E}_{n}^{\rho,(i+1)}$  as shown in (\ref{eqsm_9}), is obtained by solving the optimization problem in the variables $\hat{\rho}_{n}^{i+1}$, $ \bm{P}_{n}^{i+1}$ and nonnegative scalars $\tau^{1}\geq0$,  $\tau^{2}\geq0$,  $\tau_{l}^3\geq0$, $l = 1, \cdots, L$, $\tau_{l}^4\geq0$, $l\in\mathcal{L}^{d}$,
\begin{align}
\label{mfq-1}&\min\ \ \text{trace}(\bm{P}_{n}^{i+1})\\
&\nonumber\text{subject\ to}\ \\
\label{mfq-2}\begin{split}
&-\tau^{1}\leq0, -\tau^{2}\leq0, -\tau_{l}^3\leq0, l = 1, \cdots, L,\\  &-\tau_{l}^4\leq0,l\in\mathcal{L}^{d},\\
&\begin{bmatrix}
-\bm{P}_{n}^{i+1}&\Psi_{n}^{\rho}\\
(\Psi_{n}^{\rho})^T&-\Xi_{n}
\end{bmatrix}\preceq0,
\end{split}
\end{align}
where 
\begin{align}
\begin{split}\label{mfq-42}
\Xi_n&=diag(1,0,0)+\tau^{1}\Phi_{n}^{\rho}+\tau^{2}\Phi^s_{n}\\
&+\sum_{l=1}^{L}\tau_{l}^3\Phi_{n,l}^u+\sum_{l\in\mathcal{L}^{d}}\tau_{l}^4\Phi_{n,l}^d,
\end{split}
\end{align}
\begin{align}
\Phi_{n,l}^u&=-\begin{bmatrix}
r_l^Tr_l&-\frac{1}{2\tilde{U}_{n,l}}&-r_l^T\\
-\frac{1}{2\tilde{U}_{n,l}}&0&\bm{0}\\
-r_l&\bm{0}&\bm{I}
\end{bmatrix},\\
\label{mfq-39}\Phi_{n,l}^d&=\begin{bmatrix}
r_l^Tr_l&-\frac{1}{2\tilde{D}_{n,l}}&-r_l^T\\
-\frac{1}{2\tilde{D}_{n,l}}&0&\bm{0}\\
-r_l&\bm{0}&\bm{I}
\end{bmatrix},\\
\label{mfq-40}\Phi_{n}^{\rho}&=\begin{bmatrix}
(\hat{\rho}_{n}^i)^T(\bm{P}_{n}^i)^{-1}\hat{\rho}_{n}^i-1&\bm{0}&-(\hat{\rho}_{n}^i)^T(\bm{P}_{n}^i)^{-1}\\
\bm{0}&\bm{0}&\bm{0}\\
-((\hat{\rho}_{n}^i)^T(\bm{P}_{n}^i)^{-1})&\bm{0}&(\bm{P}_{n}^i)^{-1}
\end{bmatrix},\\
\label{mfq-41}\Phi_{n}^s&=\begin{bmatrix}
D_{n}^sU_{n}^s&-\frac{D_{n}^s+U_{n}^s}{2}&\bm{0}\\
-\frac{D_{n}^s+U_{n}^s}{2}&1&\bm{0}\\
\bm{0}&\bm{0}&\bm{0}
\end{bmatrix},\\
\Psi_{n}^{\rho}&=\begin{bmatrix}
-\hat{\rho}_{n}^{i+1}&\bm{0}&\bm{I}
\end{bmatrix}.
\end{align}
$
D_{n}^s=\min\{(\hat{s}_{n}^i-S_{n}^i)^{2/\alpha_1},(\hat{s}_{n}^i-S_{n}^i)^{2/\alpha_2}\}$, $U_{n}^s=\max\{(\hat{S}_{n}^i+s_{n}^i)^{2/\alpha_1},(\hat{s}_{n}^i+S_{n}^i)^{2/\alpha_2}\}$, \\ 
$
\tilde{D}_{n,l}=\min\{(\max\{\frac{y_l-\hat{U}_{n,l}^i}{g_l},0\})^{2/\alpha_1},(\max\{\frac{y_l-\hat{U}_{n,l}^i}{g_l},0\})^{2/\alpha_2}\}$,
$\tilde{U}_{n,l}=\min\{(\frac{y_l-\hat{D}_{n,l}^i}{g_l})^{2/\alpha_1},(\frac{y_l-\hat{D}_{n,l}^i}{g_l})^{2/\alpha_2}\}$, 
$\hat{D}_{n,l}=\sum_{j\neq n}\breve{D}_{l,j}+\bm{D}^{\varepsilon}_l$, and $\hat{U}_{n,l}=\sum_{j\neq n}\breve{U}_{l,j}+\bm{U}^{\varepsilon}_l$.
\begin{align}
\mathcal{L}^{d}=\{l:\tilde{D}_{n,l}>0,l=1,\cdots,L\}.
\end{align}

(b) The state bounding set $\mathcal{E}_{n}^{s,(i+1)}$  as shown in (\ref{eqsm_9}),  is obtained by solving the optimization problem in the variables $\tilde{s}_{n}$, $ \tilde{S}_{n}$ and nonnegative scalars $\tau^1\geq0$,  $\tau^2\geq0$,  $\tau_{l}^3\geq0$, $l = 1, \cdots, L$, $\tau_{l}^4\geq0$, $l\in\mathcal{L}^{d}$,
\begin{align}
\label{mfq-20}&\min\ \  \tilde{S}_{n}\\
&\nonumber\text{subject\ to}\ \\
\label{mfq-21}\begin{split}
&-\tau^1\leq0, -\tau^2\leq0, -\tau_{l}^3\leq0, l = 1, \cdots, L,\\  &-\tau_{l}^4\leq0,l\in\mathcal{L}^{d},\\
&\begin{bmatrix}
-\tilde{S}_{n}&\Psi_{n}^s\\
(\Psi_{n}^s)^T&-\Xi_{n}
\end{bmatrix}\preceq0,
\end{split}
\end{align}
where $
\Psi_{n}^s=\begin{bmatrix}
-\tilde{S}_{n}&1&\bm{0}
\end{bmatrix},
$
and other symbols are same as in (a).

We can obtain
\begin{align*}
\hat{S}_{n}^{i+1}=\frac{U_s+D_s}{2},~
s_{n}^{i+1}=\frac{U_s-D_s}{2},
\end{align*} 
where $D_s=\min\{(\tilde{s}_{n}^*-\tilde{S}_{n}^*)^{\alpha_1/2},(\tilde{s}_{n}^*-\tilde{S}_{n}^*)^{\alpha_2/2}\}$, $U_s=\max\{(\tilde{s}_{n}^*+\tilde{S}_{n}^*)^{\alpha_1/2},(\tilde{s}_{n}^*+\tilde{S}_{n}^*)^{\alpha_2/2}\}$ and $\tilde{s}_{n}^*$, $\tilde{S}_{n}^*$ are the optimal solution of the problem (\ref{mfq-20})-(\ref{mfq-21}).
\end{prop}
\begin{pf}
	See the Appendix.
\end{pf}

Using Lemma \ref{lemma-1} and Proposition \ref{proposition_3}, we have the alternating optimization iterative algorithm, Algorithm \ref{alg_3}, for the multiple-source localization problem with unknown but bounded $\alpha$.
\begin{algorithm}[htb]
	\caption{Ellipsoidal localization  algorithm with  unknown but bounded $\alpha$}
	\label{alg_3}
	\begin{algorithmic}[1]
			\Require
	$g(\bm{P})$: objective function,
		$\mathcal{E}_{1}^0\times$$\cdots\times\mathcal{E}_{N}^0$: initial state bounding set,
		$\mathcal{B}^{\varepsilon}$: the bounding, 
		$\delta>0$: tolerance.
		\Repeat
		\State{\textbf{Input}: the current state bound $\prod_{n=1}^N\mathcal{E}_{n}^i$.}
		\State Derive the bounding intervals $[\breve{D}_{n,l}^i,\breve{U}_{n,l}^i]$, $l=1,\cdots,L,n=1,\cdots,N$, by Lemma \ref{lemma-1}.
		\For{$n=1,\cdots,N$}
		\begin{enumerate}
			\item Optimize the center and shape matrix of the state bounding ellipsoid $\mathcal{E}_{n}^{\rho,(i+1)}$ by the problem (\ref{mfq-1})-(\ref{mfq-2}) in Proposition \ref{proposition_3}.
			\item Optimize the center and shape matrix of the state bounding ellipsoid $\mathcal{E}_{n}^{s,(i+1)}$ by the problem (\ref{mfq-20})-(\ref{mfq-21}) in Proposition \ref{proposition_3}.
		\end{enumerate}
		\EndFor
		\State \textbf{Output}: the current state bound $\mathcal{E}_{1}^{i+1}\times\cdots\times\mathcal{E}_{N}^{i+1}$.
		\Until{($g(\bm{P}^{i})-g(\bm{P}^{i+1})\leq \delta$)}
	\end{algorithmic}
\end{algorithm}

\section{Simulation  Results }\label{sec_6}

In this section, we  compare the performances of  Algorithm \ref{alg_1} (Alg-1) and Algorithm \ref{alg_2} (Alg-2) with that of the EM method in \cite{Meng-Xiao-Xie11}.  For performance comparison, we employ the  Mean Squared Error (MSE) based on $200$ Monte Carlo runs. The size of the sensor field is $100 m \times 100 m$.

We use the measurement equation (\ref{eqsm_2}) to generate the acoustic energy reading for each sensor. The gain factors for all the  sensors are equal to 1 and  the decay factor $\alpha = 2$.
The measurement noises are   independent random variables with truncated  Gaussian mixture  distribution, i.e.,
$\varepsilon_l\sim h_l(x)=c(\frac{1}{2}h_{l,1}(x)+\frac{1}{2}h_{l,2}(x))I_{[-\frac{b^{\varepsilon}(l)}{2},\frac{b^{\varepsilon}(l)}{2}]}(x), $
where $b^{\varepsilon}=\frac{\bm{U}^{\varepsilon}-\bm{D}^{\varepsilon}}{2}$ is the  vector of the side lengths of the bounding box of the measurement noise, $\bm{U}^{\varepsilon}$ and $\bm{D}^{\varepsilon}$ are defined in (\ref{eqsm_5}). $h_{l,i}\sim\mathcal{N}(\mu_i,\sigma_{i}^2),i=1,2$, $\mu_1=\frac{b^{\varepsilon}_l}{2}$, $\sigma_{1}=\frac{b^{\varepsilon}_l}{6}$, $\mu_2=-\frac{b^{\varepsilon}_l}{2}$, $\sigma_{2}=\frac{b^{\varepsilon}_l}{6}$, $c$ is the normalizing constant, and  $I_{[-\frac{b^{\varepsilon}_l}{2},\frac{b^{\varepsilon}_l}{2}]}(x)$ is an indicator function of the set $[-\frac{b^{\varepsilon}_l}{2},\frac{b^{\varepsilon}_l}{2}]$. Moreover, set all the components of the vector $b^{\varepsilon}$ to be equal.
\begin{figure}
	\begin{center}
		\includegraphics[height=6cm]{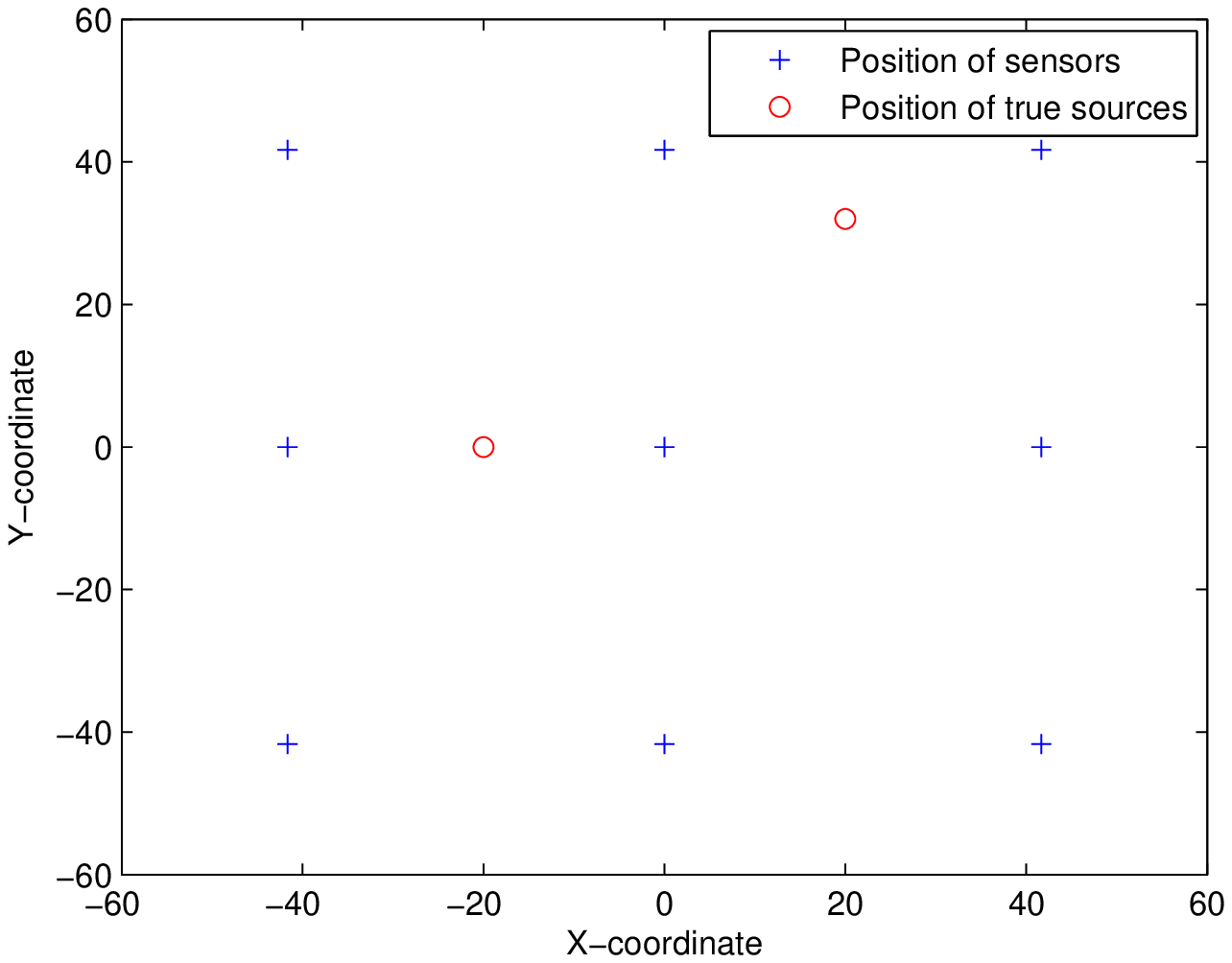}    
		\caption{The positions of two sources and the sensors placed on a grid.}  
		\label{fig_Sim_01}                                 
	\end{center}                                 
\end{figure}
\begin{figure}
	\begin{center}
		\includegraphics[height=6cm]{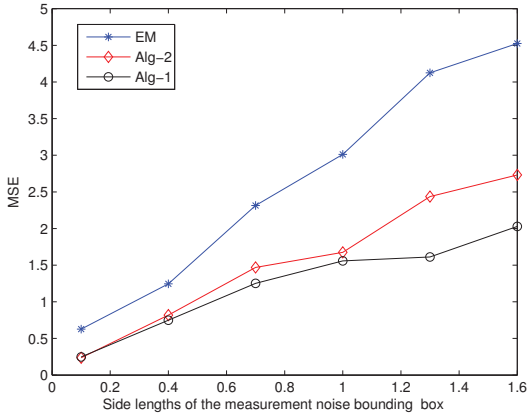}    
		\caption{MSE of the state estimates for two sources and  sensor locations are shown in Fig. \ref{fig_Sim_01}  is plotted as a function of  $b^{\varepsilon}$.}  
		\label{fig_Sim_02}                                 
	\end{center}                                 
\end{figure}

Nine sensor nodes are placed on a  regular grid as shown in Fig. \ref{fig_Sim_01}. The sensor locations remain fixed during all the Monte Carlo runs. Two acoustic sources are located at $[-20, 0]$ and $[20, 32]$, respectively.  The  source energies are assumed to be  $s_1=6000$, and  $s_2 = 6500$. The initial state bounding set  is    $\mathcal{E}^0$, shown in (\ref{eqsm_8}). For the $n$-th source, $n=1$, $2$, the Cartesian product of  ellipsoids $\mathcal{E}_{n}^0=\mathcal{E}_{n}^{s,0}\times\mathcal{E}_{n}^{\rho,0}$  is randomly selected in each run, where $\mathcal{E}_{n}^{s,0}$ is an interval of length 200 and $\mathcal{E}_{n}^{\rho,0}$ is a ball of radius  $7m$. The MSEs of Alg-1, Alg-2 and the EM method are plotted as a function of $b^{\varepsilon}$ in Fig. \ref{fig_Sim_02}, respectively. 
To further understand the simplified Algorithm \ref{alg_2}, we have  presented   additional  numerical simulations, see Fig. \ref{fig_Sim_11}-Fig. \ref{fig_Sim_14}. The state estimates of the  two sources are plotted as a function of the number of iterations, number of sensors and distance between two sources, respectively.

\begin{figure}
	\begin{center}
		\includegraphics[height=6cm]{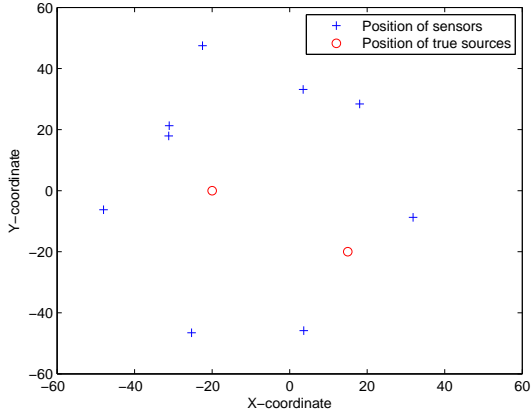}    
		\caption{The positions of the sources and the sensors when sensors are located  randomly in the region of interest.}  
		\label{fig_Sim_05}                                 
	\end{center}                                 
\end{figure}

\begin{figure}
	\begin{center}
		\includegraphics[height=6cm]{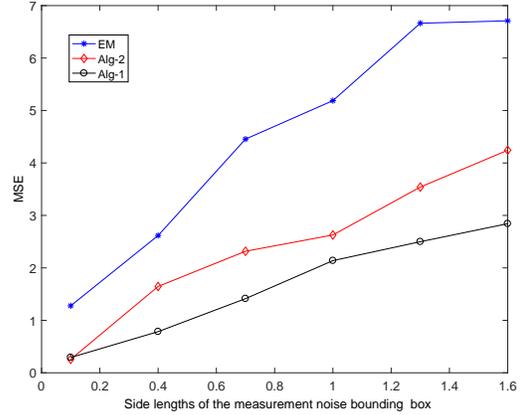}    
		\caption{MSE of the state estimates for two sources for randomly located sensors as shown in  Fig. \ref{fig_Sim_05}  is plotted as a function of  $b^{\varepsilon}$.}  
		\label{fig_Sim_06}                                 
	\end{center}                                 
\end{figure}

\begin{figure}
	\begin{center}
		\includegraphics[height=6cm]{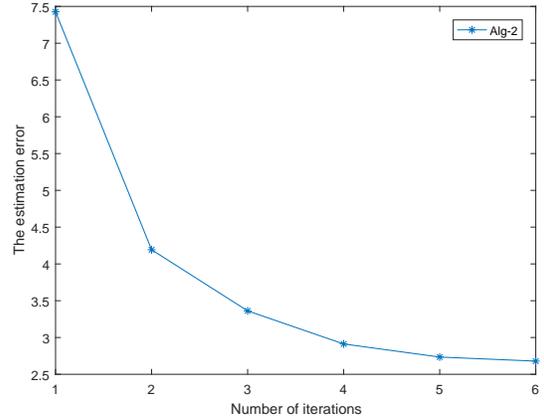}    
		\caption{The state estimation error for two sources  is plotted as a function of the number of  iterations.}  
		\label{fig_Sim_11}                                 
	\end{center}                                 
\end{figure}

\begin{figure}
	\begin{center}
		\includegraphics[height=6cm]{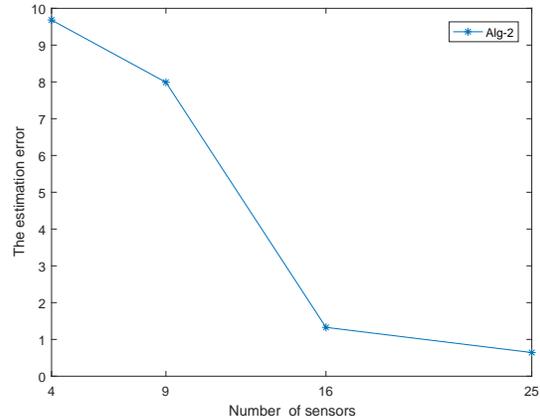}    
		\caption{The state estimation error for two sources  is plotted as a function of the number of sensors.}  
		\label{fig_Sim_12}                                 
	\end{center}                                 
\end{figure}

\begin{figure}
	\begin{center}
		\includegraphics[height=6cm]{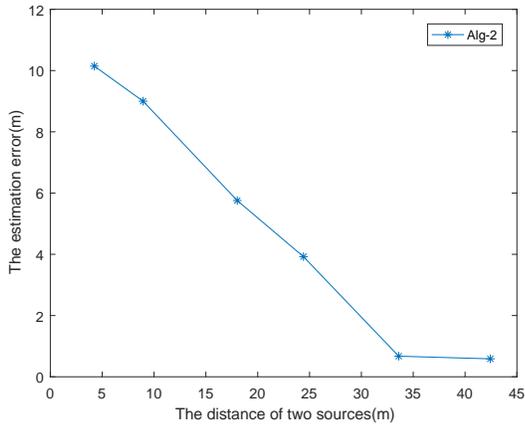}    
		\caption{The state estimation error for two sources is plotted as a function of the distance between two sources.}  
		\label{fig_Sim_14}                                 
	\end{center}                                 
\end{figure}

Moreover, we also consider a scenario in which  sensor node  locations  are not on a grid and are random   as shown in Fig. \ref{fig_Sim_05}.  Two acoustic sources are located at $[-20, 0]$ and $[15, -20]$, respectively. The MSEs of Alg-1, Alg-2 and the EM method are plotted as a function of $b_{\varepsilon}$ in Fig. \ref{fig_Sim_06}, respectively.


\begin{figure}
	\begin{center}
		\includegraphics[height=6cm]{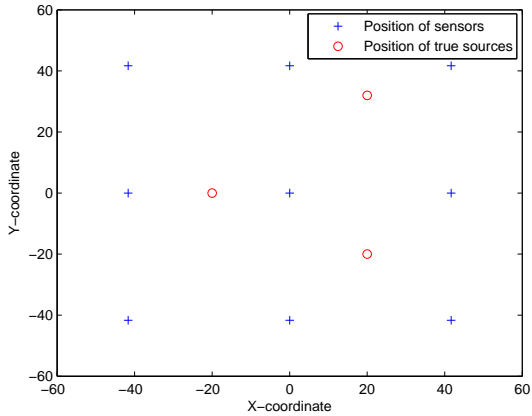}    
		\caption{The positions of three sources and the sensors placed on a grid.}  
		\label{fig_Sim_07}                                 
	\end{center}                                 
\end{figure}

\begin{figure}
	\begin{center}
		\includegraphics[height=6cm]{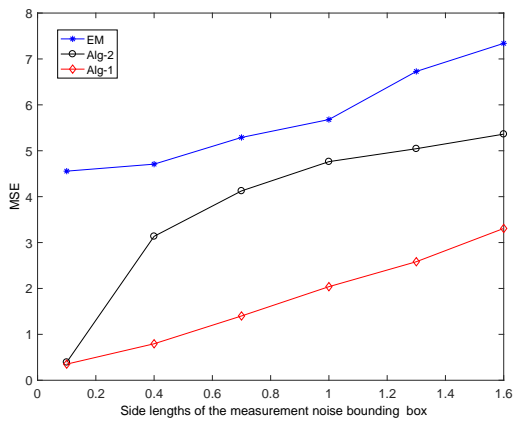}    
		\caption{MSE of the state estimates for three sources with sensors located on a grid as shown in Fig. \ref{fig_Sim_07}   is plotted as a function of  $b^{\varepsilon}$.}  
		\label{fig_Sim_08}                                 
	\end{center}                                 
\end{figure}

\begin{figure}
	\begin{center}
		\includegraphics[height=6cm]{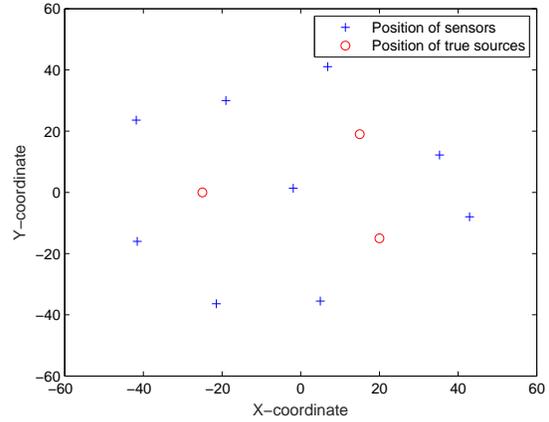}    
		\caption{The positions of three sources and the sensors placed randomly in the region of interest.}  
		\label{fig_Sim_09}                                 
	\end{center}                                 
\end{figure}

\begin{figure}
	\begin{center}
		\includegraphics[height=6cm]{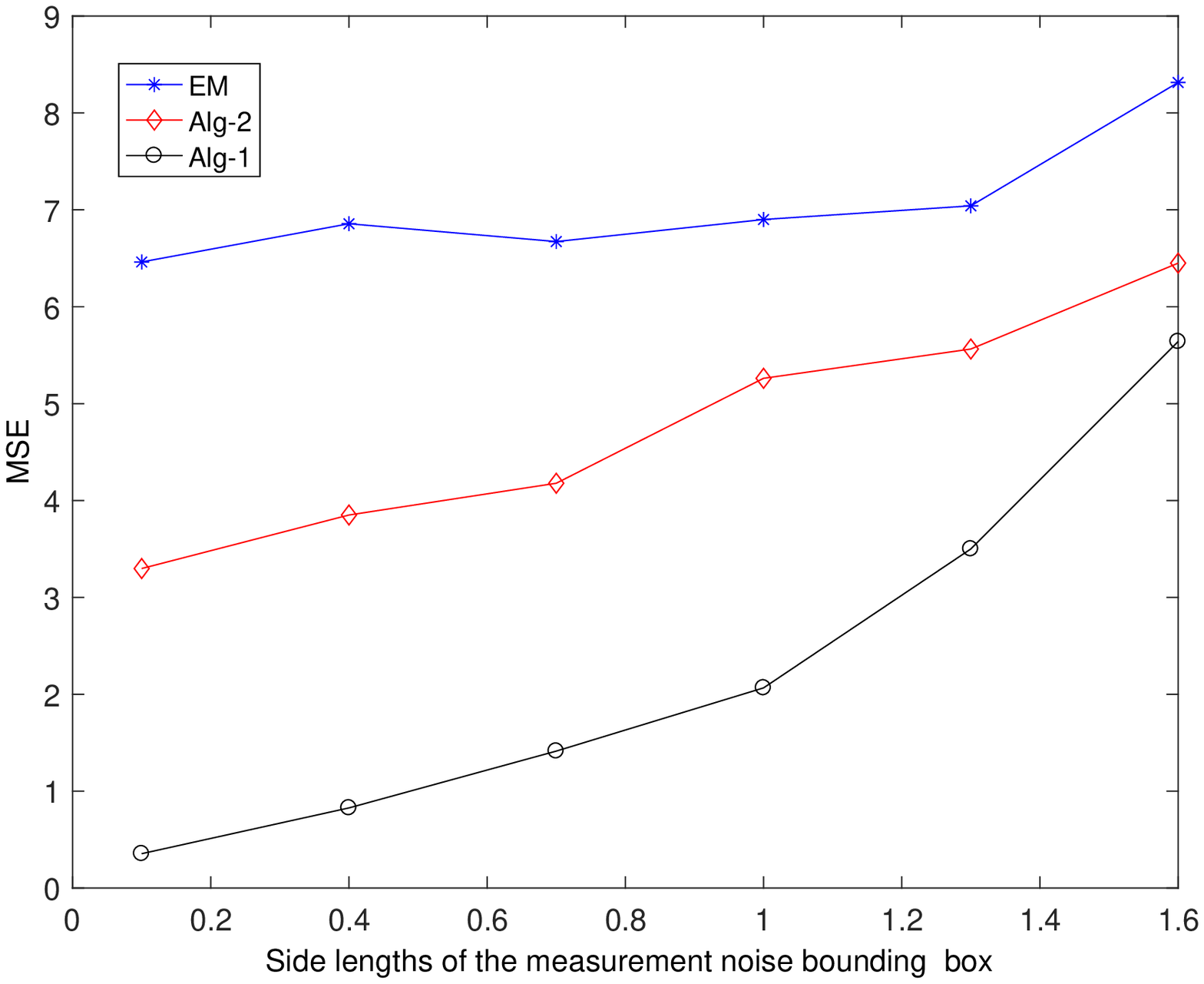}    
		\caption{MSE of the state estimates for tree sources  with sensors located as shown in Fig. \ref{fig_Sim_09}  is plotted as a function of  $b^{\varepsilon}$.}  
		\label{fig_Sim_10}                                 
	\end{center}                                 
\end{figure}

Similarly, the  localization problem for three sensors  is considered next. Sensor nodes are placed as shown in Fig.\ref{fig_Sim_07} and Fig.\ref{fig_Sim_09}, respectively. As  shown in Fig.\ref{fig_Sim_07}, three acoustic sources are located at $[-20, 0]$, $[20, 32]$ and $[20,-20]$, respectively. The source energies are $s_1=6000$,  $s_2 = 6500$ and  $s_3 = 6800$ respectively. The MSEs of Alg-1, Alg-2 and the EM method are plotted as a function of $b_{\varepsilon}$ in Fig. \ref{fig_Sim_08}. The locations of all the  sensors in Fig.\ref{fig_Sim_09} are random. Three acoustic sources are located at $[-25, 0]$, $[15, 19]$ and $[20, -15]$, respectively. The MSEs of Alg-1, Alg-2 and the EM method are plotted as a function of $b^{\varepsilon}$ in Fig. \ref{fig_Sim_10}. The computation  times of the three algorithms are shown in Table \ref{tab:1}. The  time in  each case is the mean of  200 Monte Carlo runs.

\begin{table}[http]
\begin{center}
\caption{The Computation  Time of Different Algorithms }\label{tab:1}
	\begin{tabular}{|c|c|c|c|}
		\hline
		\diagbox{Source}{time(s)}{method}&EM&Alg-1&Alg-2\\
		\hline
		two sources&  3.03&46.3 &24.03\\
		\hline
		three sources &16.8&76.6&52.2\\
		\hline
	\end{tabular}
	
\end{center}
\end{table}

\begin{figure}
	\begin{center}
		\includegraphics[height=6cm]{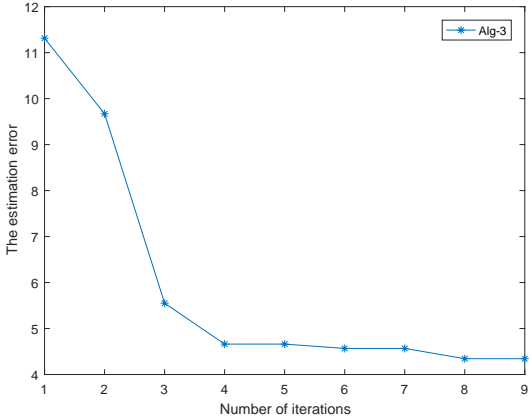}    
		\caption{The state estimation error for two sources when $\alpha$ is unknown but bounded, is plotted as a function of the  number of iterations.}  
		\label{fig_Sim_15}                                 
	\end{center}                                 
\end{figure}

When the energy decay factor $\alpha$ is unknown but bounded, i.e., $\alpha\in[2.8,3.2]$, the estimation error of  Algorithm \ref{alg_3} (Alg-\ref{alg_3}) is shown in Fig. \ref{fig_Sim_15} where there are two sources and 16 sensors. It is plotted as a function of the number of iterations.

From Figs. \ref{fig_Sim_01}-\ref{fig_Sim_15}, we make the following observations:
\begin{itemize}
	\item Figs. \ref{fig_Sim_01}-\ref{fig_Sim_05} and Figs. \ref{fig_Sim_07}-\ref{fig_Sim_10} show that the performances of both Alg-1 and Alg-2 are better than that of the EM method.  The main  reason is that the EM method is based on the Gaussian assumption. However, in this example, the measurement noise is non-Gaussian. Actually, the ellipsoidal localization approach presented in this paper in the unknown but bounded setting only depends on the bounds of noises and does not  rely  on the probability density function (PDF). In addition, the figures show that the larger is the noise bound $b^{\varepsilon}$, the larger is the MSE.
	\item Figs. \ref{fig_Sim_01}-\ref{fig_Sim_05} and Figs. \ref{fig_Sim_07}-\ref{fig_Sim_10} also show that the performance of Alg-1 is  better than that of Alg-2 since the ellipsoidal state  bound $\mathcal{E}^{\rho}$ in Proposition \ref{proposition_1_1} is relaxed to a bounding ball in Proposition \ref{proposition_1} where the radius is the long semi-axis of the ellipsoid $\mathcal{E}^{\rho}$ . However,  Alg-2  requires less computation time than Alg-1, as shown in Table \ref{tab:1}. The reason is that the  solution of the remainder bound is obtained by solving an SDP problem in  Proposition \ref{proposition_1_1} whereas  the bounding box of the remainder is obtained analytically in  Proposition \ref{proposition_1}. Thus, there is a tradeoff between computation time and localization accuracy.
	\item The estimation error of Alg-2 is plotted as a function of  the number of iterations in Fig. \ref{fig_Sim_11}. It shows that the performance improves  with the number of iterations and then it becomes stable. Fig. \ref{fig_Sim_12} shows that the estimation improves with the number of sensors.   Fig. \ref{fig_Sim_14} shows that the distance between the  sources  affects the localization accuracy. If  sources are too close, the accuracy decreases. With an increase of source spacing, the localization accuracy becomes better.  Fig. \ref{fig_Sim_15}  shows that  Alg-\ref{alg_3} can deal with the multiple source localization problem when $\alpha$ is unknown but bounded. The estimation error decreases with the number of iterations.
	\item As shown in Figs. 6 and 13, the  estimation accuracy improves  slowly after several iterations. The estimation accuracy increases with the decrease of  $\delta$, but the  computation  time  increases at the same time. 
	For the trade-off between the computation  time and the  estimation accuracy, a reasonable value for the  termination  criterion might be $\delta=10^{-2}$.  Furthermore, the numerical simulation also shows that the size of the final shape matrix becomes stable with the number of iterations. When the  value of $\delta$ is smaller than $10^{-2}$, the size of the final shape matrix does not change much. 
	\item In summary, numerical examples show that when the PDF of measurement noise is unknown-but-bounded,  Alg-1 is  most effective for multiple-source localization as far as the estimation performance is considered. The computation time of the EM algorithm is the smallest.  Alg-2  provides  a good trade off between  estimation performance and computation time.
\end{itemize}

\section{Conclusion}\label{sec_7}
In this paper, we have proposed  new multiple-source localization methods in the unknown-but-bounded noise setting. We employed  set-membership estimation theory to determine a state estimation ellipsoid.
The main difficulties are that the acoustic energy decay model  is a complicated nonlinear function and the multiple-source localization problem is a high-dimensional state estimation problem. 
In our approach,  the nonlinear function is linearized by the first-order Taylor series expansion with a remainder error. The bounding box of the remainder has been derived on-line based on the bounding set of the state. A point that should be stressed is that the remainder bounding box is obtained  analytically when the energy parameter and the position of the source are bounded
in an interval and a ball  respectively.
An efficient procedure has been developed to deal with the multiple-source
localization problem by alternately estimating the parameters of each source while the parameters of the other sources remain  fixed. When the energy decay factor is unknown but bounded, a new estimation procedure has been developed. 
Numerical examples have shown that when the PDF of measurement noise is non-Gaussian, the performance of the ellipsoidal localization algorithm  is better than the ML method.
Future work may include   sensor management and sensor placement,  Byzantines
and mitigation techniques, for the multiple-source localization problem  in wireless sensor networks.

\appendix
\section{Proof of Proposition \ref{proposition_1_1}}    
 Let  $\bm{x}_n=\hat{\bm{x}}_n^i+[\Delta s,\Delta\rho^T]^T$,  the function $\Delta f_{n,l}(\bm{x}_n,\hat{\bm{x}}_n^i)$ is rewritten as 
\begin{align}\label{Ap_1}
\begin{split}
\Delta \tilde{f}_{n,l}&(\Delta\rho,\Delta s)
=g_l\bigg(\frac{\hat{s}_n^i+\Delta s}{\|\hat{\rho}_n^i-r_l+\Delta\rho\|^\alpha}\\
&-\frac{\hat{s}_n^i+\Delta s}{\|\hat{\rho}_n^i-r_l\|^\alpha}+\alpha\hat{s}_n^i\frac{(\hat{\rho}_n^i-r_l)^T}{ \|\hat{\rho}_n^i-r_l\|^{\alpha+2}}\Delta\rho\bigg).
\end{split}
\end{align}
The derived functions of $\Delta \tilde{f}_{n,l}$ are
\begin{align*}
&\frac{\partial\Delta \tilde{f}_{n,l}(\Delta\rho,\Delta s)}{\partial \Delta\rho}
=g_l\bigg(\frac{\alpha \hat{s}_n^i}{\|\hat{\rho}_n^i-r_l\|^{\alpha+2}}(\hat{\rho}_n^i-r_l)
\\
&~~~~~~~~~~~~~~~~-\frac{\alpha(\hat{s}_n^i+\Delta s)}{\|\hat{\rho}_n^i-r_l+\Delta\rho\|^{\alpha+2}}(\hat{\rho}_n^i-r_l+\Delta\rho)
\bigg),\\
&\frac{\partial\Delta \tilde{f}_{n,l}(\Delta\rho,\Delta s)}{\partial \Delta s}
=g_l\bigg(
\frac{1}{\|\hat{\rho}_n^i-r_l+\Delta\rho\|^{\alpha}}\\
&~~~~~~~~~~~~~~~~-\frac{1}{\|\hat{\rho}_n^i-r_l\|^\alpha}
\bigg).
\end{align*}
If $[\Delta s,\Delta\rho^T]^T$ is a stationary point, then
$$\frac{\partial\Delta \tilde{f}_{n,l}(\Delta\rho,\Delta s)}{\partial \Delta\rho}=0,\frac{\partial\Delta \tilde{f}_{n,l}(\Delta\rho,\Delta s)}{\partial \Delta s}=0.$$
We can get
\begin{align*}
\frac{|\hat{s}_n^i+\Delta s|}{\hat{s}_n^i}\|\hat{\rho}_n^i-r_l+\Delta\rho\|&=\|\hat{\rho}_n^i-r_l\|,\\
\|\hat{\rho}_n^i-r_l+\Delta\rho\|&=\|\hat{\rho}_n^i-r_l\|,
\end{align*}
where $\|\hat{\rho}_n^i-r_l\|>0,\ \hat{s}_n^i+\Delta s>0$, then $\Delta s=0$  and  $\Delta\rho=0$.  It meas that there is only one stationary point in $\mathcal{E}^i_n$ and it is $\hat{\bm{x}}_n^i=[\hat{s}_n^i,(\hat{\rho}_n^i)^T]^T$.

If the $l$-th sensor is not contained in the set $\mathcal{E}_{n}^{\rho,i}$, then the function $\Delta f_{n,l}(\bm{x}_n,\hat{\bm{x}}_{n}^i)$ is differentiable on the set  $\mathcal{E}_{n}^i$. 
Since there is only one stationary point in $\mathcal{E}_n^i$, the minimum and maximum of $\Delta f_{n,l}(\bm{x}_n,\hat{\bm{x}}_{n}^i)$ are  obtained at the stationary point $\hat{\bm{x}}_n^i$ or  on the boundary of $\mathcal{E}_n^{i}$. 

If the $l$-th sensor is contained in the set $\mathcal{E}_{n}^{\rho,i}$,  for the same reason, then the minimum of $\Delta f_{n,l}(\bm{x}_n,\hat{\bm{x}}_{n}^i)$ is  obtained at the stationary point $\hat{\bm{x}}_n^i$ or on the boundary of $\mathcal{E}_n^{i}$. The maximum of $\Delta f_{n,l}(\bm{x}_n,\hat{\bm{x}}_{n}^i)$  is $+\infty$ and it is obtained at  $\Delta\rho=-(\hat{\rho}_n^i-r_l)$. 

Therefore, we have Proposition \ref{proposition_1_1}.
$\Box$
\section{Proof of Proposition \ref{proposition_1}}
For $l=1,\cdots,L$,
$\Delta \tilde{f}_{n,l}(\Delta\rho,\Delta s)$ is defined  (\ref{Ap_1}).
Since $g_l$ is a positive constant, we only consider the following function
\begin{align}\label{Corollary_1_eq_2}
\begin{split}
&H(\Delta\rho,\Delta S)=\frac{\Delta \tilde{f}_{n,l}(\Delta\rho,\Delta s)}{g_l}.
\end{split}
\end{align}
Let $\tau_l=\|\hat{\rho}^i_n-r_l\|$, $t=\|\Delta\rho\|\in[0,R_n^i]$, $k=\cos(\theta)$ and $\theta\in[0,\pi]$, so that we can rewrite  (\ref{Corollary_1_eq_2})  as
\begin{align*}
H(k,\Delta s,t)&=\frac{\hat{s}_n^i+\Delta s}{(t^2+\tau_l^2+2t\tau_l k)^{\alpha/2}}+\frac{\alpha\hat{s}_n^itk}{\tau_l^{\alpha+1}}-\frac{\hat{s}_n^i+\Delta s}{\tau_l^\alpha}, 
\end{align*}
where $k\in[-1,1]$, $\Delta s\in[-S_n^i,S_n^i]$, $t\in[0,R_n^i]$,  and $R_n^i<\tau_l$.
%
%

The proof falls naturally  into two parts: ($\uppercase\expandafter{\romannumeral1}$) find the maximum of the function $H(k,\Delta s,t)$ when $k\in[-1,1],\ \Delta s\in[-S_n^i,S_n^i],\ t\in[0,R_n^i]$, ($\uppercase\expandafter{\romannumeral2}$) find the minimum of the function $H(k,\Delta s,t)$ when $k\in[-1,1],\ \Delta s\in[-S_n^i,S_n^i],\ \ t\in[0,R_n^i]$.



($\uppercase\expandafter{\romannumeral1}$) It is clear that $H(k,\Delta s,t)$ is a linear function of $\Delta s$. 
Moreover, by Proposition \ref{proposition_1_1}, in order to get  the maximum  of the function $H(k,\Delta s,t)$ on $k$ and $t$, we only need to consider the two cases:  ($\romannumeral1$) $k\in[-1,1],\ \Delta s=-S_n^i,\ t\in[0,R_n^i]$, ($\romannumeral2$) $k\in[-1,1],\ \Delta s=S_n^i,\ t\in[0,R_n^i]$.

($\romannumeral1$) Since
$H(k,-S_n^i,t)$ is a convex function of $k$, the maximum is obtained at $k=1\ or\ -1$. It means that we only need to consider two functions of $t$:
 $H(1,-S_n^i,t)$ and $H(-1,-S_n^i,t)$.

Since $H(1,-S_n^i,t)$ is a convex function of $t$, the maximum is  obtained at $t = 0$ or $R_n^i$. It is easily seen that $H(1,-S_n^i,0)=0$. We get 
\begin{align}
\max_{t\in[0,R_n^i]}H(1,-S_n^i,t)&=\max\{H(1,-S_n^i,R_n^i),0\}\label{Corollary_1_eq_9}.
\end{align}
In the same manner we can see that 
%
%
\begin{align}
\max_{t\in[0,R_n^i]}H(-1,-S_n^i,t)&=\max\{H(-1,-S_n^i,R_n^i),0\}\label{Corollary_1_eq_11}.
\end{align}
($\romannumeral2$) 
For 
$H(k,S_n^i,t)$ is a convex function of $k$,  the maximum is obtained at $k=1\ or\ -1$.
It means that we only need to consider two functions of $t$:  $H(1,S_n^i,t)$ and $H(-1,S_n^i,t)$.

It is obvious that $H(1,\Delta s,t)$ is a monotonic decreasing function of $\Delta s$. We  have  $H(1,S_n^i,t)\leq H(1,-S_n^i,t)$ and
\begin{align}\label{Corollary_1_eq_43}
\max_{t\in[0,R_n^i]}H(1,S_n^i,t)\leq \max_{t\in[0,R_n^i]}H(1,-S_n^i,t).
\end{align}
Since $H(-1,S_n^i,t)$ is a convex function of $t$, the maximum is  obtained at $t = 0$ or $R_n^i$. We have $H(1,S_n^i,0)=0$ and
\begin{align}
\max_{t\in[0,R_n^i]}H(-1,S_n^i,t)&=\max\{H(-1,S_n^i,R_n^i),0\}\label{Corollary_1_eq_18}.
\end{align}

Combining (\ref{Corollary_1_eq_9}),    (\ref{Corollary_1_eq_11}),   (\ref{Corollary_1_eq_43}) and   (\ref{Corollary_1_eq_18}), we conclude that 
\begin{align}\label{Corollary_1_eq_19}
\begin{split}
&\max_{k\in[-1,1], \Delta s\in[-S_n^i,S_n^i],t\in[0,R_n^i]}H(k,\Delta s,t)\\
&=\max\{H(1,-S_n^i,R_n^i),H(-1,S_n^i,R_n^i),\\
&~~~~~~~~~~~~~~~~~~~~~~~H(-1,-S_n^i,R_n^i),0\}.
\end{split}
\end{align}

($\uppercase\expandafter{\romannumeral2}$) Similar arguments apply to this case, in order to get  the minimum  of the function $H(k,\Delta s,t)$ on $k$ and $t$, we only need to consider the two cases:  ($\romannumeral1$) $k\in[-1,1],\ \Delta s=-S_n^i,\ t\in[0,R_n^i]$, ($\romannumeral2$) $k\in[-1,1],\ \Delta s=S_n^i,\ t\in[0,R_n^i]$.

($\romannumeral1$) Let $\frac{\partial H(k,-S_n^i,t)}{\partial k}=0$,
we have
\begin{align}\label{Corollary_1_eq_21}
\hat{k}_1=\frac{\bigg((1-\frac{S_n^i}{\hat{s}_n^i})^{1/(\alpha/2+1)}-1\bigg)\tau_l^2-t^2}{2t\tau_l}\leq0.
\end{align}
Since $H(k,-S_n^i,t)$ is a convex function of $k$, the minimum is obtained at $k=\hat{k}_1$ when $\hat{k}_1\in[-1,1]$. If $\hat{k}_1\notin[-1,1]$, the minimum is obtained at $k=-1$ when $\hat{k}_1\notin[-1,1]$.

Let $\frac{\partial H(-1,-S_n^i,t)}{\partial t}=0$, 
we get
\begin{align*}
t=\tau\bigg(1-(1-\frac{S_n^i}{\hat{s}_n^i})^{1/(\alpha+1)}\bigg)\geq0.
\end{align*} 
Since $H(-1,-S_n^i,t)$ is a convex function of $t$, we obtain
\begin{align}\label{Corollary_1_eq_27}
\min_{t\in[0,R_n^i]}H(-1,-S_n^i,t)=H(-1,-S_n^i,t_1).
\end{align}
where $t_1$ is defined in (\ref{mfq-23}).

Consider the function
\begin{align*}
H(\hat{k},-S_n^i,t)=\hat{s}_n^i\bigg(&\frac{2(1-\frac{S_n^i}{\hat{s}_n^i})^{1/(\alpha/2+1)}-1-(1-\frac{S_n^i}{\hat{s}_n^i})}{\tau_l^\alpha}\\
&-\frac{t^2}{\tau_l^{\alpha+2}}\bigg),
\end{align*}
where
$
\hat{k}_1\in[-1,1]. 
$
It is equivalent to
\begin{align}\label{Corollary_1_eq_30}
\bigg|(1-\frac{S_n^i}{\hat{s}_n^i})^{1/(\alpha+2)}-1\bigg|\tau_l\leq t.
\end{align}
In this case, the minimum  is $H(\hat{k}_1,-S_n^i,R_n^i)$.

Now we have 
\begin{align}
\begin{split}\label{mfq_51}
&\min_{k\in[-1,1],\  t\in[0,R_n^i]}H(k,-S_n^i,R_n^i)\\
&=\min\{H(-1,-S_n^i,t_1),H(\max\{\hat{k}_1,-1\},-S_n^i,R_n^i)\}.
\end{split}
\end{align}

($\romannumeral2$)
Let $\frac{\partial H(k,S_n^i,t)}{\partial k}=0,$
we have
\begin{align}\label{Corollary_1_eq_32}
\hat{k}_2=\frac{\bigg((1+\frac{S_n^i}{\hat{s}_n^i})^{1/(\alpha/2+1)}-1\bigg)\tau_l^2-t^2}{2t\tau_l}.
\end{align}
Since $H(k,S_n^i,t)$ is a convex function of $k$, the minimum is obtained at $k=\hat{k}_2$ when $\hat{k}\in[-1,1]$ and the minimum is obtained at $k=1\ or\ -1$ when $\hat{k}_2\notin[-1,1]$.

Let  $\frac{\partial H(1,S_n^i,t)}{\partial t}=0$, 
we get $t=\tau_l\bigg((1+\frac{S_n^i}{\hat{s}_n^i})^{1/(\alpha+1)}-1\bigg)>0$. 
Since $H(1,S_n^i,t)$ is a convex function of $t$,  we have
\begin{align}\label{Corollary_1_eq_35}
\min_{t\in[0,R_n^i]}H(1,S_n^i,t)=H(1,S_n^i,t_2),
\end{align}
where  $t_2$ is defined in (\ref{mfq-24}). 

Let $\frac{\partial H(-1,S_n^i,t)}{\partial t}=0$, 
we get
\begin{align*}
t=\tau\bigg(1-(1+\frac{S_n^i}{\hat{s}_n^i})^{1/(\alpha+1)}\bigg)\leq0.
\end{align*}  Since $H(-1,s,t)$ is a convex function of $t$, we obtain
\begin{align}\label{Corollary_1_eq_38}
\min_{t\in[0,R_n^i]}H(-1,S_n^i,t)=0.
\end{align}

Consider the function
\begin{align*}
H(\hat{k}_2,S_n^i,t)=\hat{s}_n^i\bigg(&\frac{2(1+\frac{S_n^i}{\hat{s}_n^i})^{1/(\alpha/2+1)}-1-(1+\frac{S_n^i}{\hat{s}_n^i})}{\tau_l^\alpha}\\
&-\frac{t^2}{\tau_l^{\alpha+2}}\bigg), 
\end{align*}
where $\hat{k}_2\in[-1,1]$.  It is equivalent to
\begin{align}\label{Corollary_1_eq_41}
\bigg|(1+\frac{S_n^i}{\hat{s}_n^i})^{1/(\alpha+2)}-1\bigg|\tau_l\leq t.
\end{align}
In this case, the minimum  is $H(\hat{k}_2,S_n^i,R_n^i)$.

Now we have 
\begin{align}
\begin{split}\label{mfq_52}
&\min_{k\in[-1,1],\  t\in[0,R_n^i]}H(k,S_n^i,R_n^i)\\
&=\min\{H(\max\{-1,\min\{\hat{k}_1,1\}\},S_n^i,R_n^i),0,\\
&~~~~~~~~~~~~~~~~~~H(1,S_n^i,t_2)\}.
\end{split}
\end{align}

Therefore, from
(\ref{mfq_51}) and (\ref{mfq_52}), we have the following result
\begin{align}\label{Corollary_1_eq_42}
\begin{split}
&\min_{k\in[-1,1], \Delta s\in[-S_n^i,S_n^i],t\in[0,R_n^i]}H(k,\Delta s,t)\\
&=\min\{H(-1,-S_n^i,t_1),H(\max\{\hat{k}_1,-1\},-S_n^i,R_n^i),\\
&\ \  H(1,S_n^i,t_2),0,H(\max\{-1,\min\{\hat{k}_1,1\}\},S_n^i,R_n^i)\}.
\end{split}
\end{align}

Base on Equations (\ref{Corollary_1_eq_2}), (\ref{Corollary_1_eq_19}) and (\ref{Corollary_1_eq_42}), we  have actually proved the Proposition \ref{proposition_1}.\ ~~~ $\Box$

\section{Proof of Proposition \ref{proposition_1_c}}
All symbols are same as those in the proof of Proposition \ref{proposition_1}.
The proof is divided  into two parts: ($\uppercase\expandafter{\romannumeral1}$) find the minimum of the function $H(k,\Delta s,t)$ when $k\in[-1,1],\ \Delta s\in[-S_n^i,S_n^i],\ t\in[0,\tau_l]$, ($\uppercase\expandafter{\romannumeral2}$) find the minimum of the function $H(k,\Delta s,t)$ when $k\in[-1,1],\ \Delta s\in[-S_n^i,S_n^i],\ \ t\in[\tau_l,R_n^i]$.

($\uppercase\expandafter{\romannumeral1}$)
Firstly, let us consider the function $H(k,\Delta s,t)$ when $k\in[-1,1],\ \Delta s\in[-S_n^i,S_n^i],\ t\in[0,\bar{\tau}_l]$, $0<\bar{\tau}_l<\tau_l$. 
From Proposition \ref{proposition_1}, we get 
\begin{align}
\begin{split}
&\min_{k\in[-1,1], \Delta s\in[-S_n^i,S_n^i],t\in[0,\bar{\tau}_l]}H(k,\Delta s,t)\\
&=\min\bigg\{H(-1,-S_n^i,t_1),H(1,S_n^i,t_2), \\
& ~~~~~~~~H(\max\{-1,\hat{k}_1\},-S_n^i,\bar{\tau}_l),\\
& ~~~~~~~~H(\max\{-1,\min\{\hat{k}_2,1\}\},S_n^i,\bar{\tau}_l),0\bigg\},
\end{split}
\end{align}
where $t_1$, $t_2$, $\hat{k}_1$ and $\hat{k}_2$ are obtained in  (\ref{mfq-23})-(\ref{mfq-26}). 

Since $H(k,\Delta s,t)=\infty$ and $H$ is continuous when $(k,t)\neq(-1,\tau_l)$, we have 
\begin{align*}
&\min_{k\in[-1,1], \Delta s\in[-S_n^i,S_n^i],t\in[0,\tau_l]}H(k,\Delta s,t)\\
&=\lim_{\tau_{l}^-\rightarrow(\tau_l)^-}\min_{k\in[-1,1], \Delta s\in[-S_n^i,S_n^i],t\in[0,\bar{\tau}_l]}H(k,\Delta s,t).
\end{align*}
Moreover, denote
\begin{align*}
t_1^-&=\lim_{\bar{\tau}_l\rightarrow\tau_l^-}t_1,~~
t_2^-=\lim_{\bar{\tau}_l\rightarrow\tau_l^-}t_2,\\
\hat{k}_1^-&=\lim_{\bar{\tau}_l\rightarrow\tau_l^-}\hat{k}_1,~~
\hat{k}_2^-=\lim_{\bar{\tau}_l\rightarrow\tau_l^-}\hat{k}_2.
\end{align*}
It is easy to find that $\hat{k}_1^->-1$ and $-1<\hat{k}_2^-<1$.
Thus, the minimum of $H(k,\Delta s,t)$  for the case of  ($\uppercase\expandafter{\romannumeral1}$) is
\begin{align}\label{result_1_c}
\begin{split}
&\min_{k\in[-1,1], \Delta s\in[-S_n^i,S_n^i],t\in[0,\tau_l]}H(k,\Delta s,t)\\
&=\min\bigg\{H(-1,-S_n^i,t_1^-), H(1,S_n^i,t_2^-),\\
& ~~~~~~~~H(\hat{k}_1^-,-S_n^i,\tau_l),H(\hat{k}_2^-,S_n^i,\tau_l),0\bigg\}.
\end{split}
\end{align}

($\uppercase\expandafter{\romannumeral2}$) 
Consider the function $H(k,\Delta s,t)$ when $k\in[-1,1],\ \Delta s\in[-S_n^i,S_n^i],\ t\in[\tau^{+},R]$, $\tau^{+}\in(\tau_l,R_n^i)$. 
Since the function $H(k,\Delta s,t)$ is a linear function of $\Delta s$,  the minimum  is obtained at $\Delta s=-S_n^i $ or $\Delta s=S_n^i$.
Moreover, by Proposition \ref{proposition_1_1},  to obtain the minimum of the function $H(k,\Delta s,t)$ on $k$ and $t$, we only need to consider the two cases: ($\romannumeral1$) $k\in[-1,1],\ \Delta s=-S_n^i,\ t\in[\tau^{+},R_n^i]$, ($\romannumeral2$) $k\in[-1,1],\ \Delta s=S_n^i,\ t\in[\tau^{+},R_n^i]$.

($\romannumeral1$) 
Let
$
\frac{\partial H(k,-S_n^i,t)}{\partial k}=0,
$ 
we get $\hat{k}_1\leq0$ (see in (\ref{Corollary_1_eq_21})).
Moreover, $\hat{k}_1\in[-1,1]$
is equivalent to (\ref{Corollary_1_eq_30}).
Since $t\geq\tau^{+}>\tau$ and $\bigg|(1-\frac{S_n^i}{\hat{s}_n^i})^{1/(\alpha+2)}-1\bigg|\leq1$, the inequality (\ref{Corollary_1_eq_30}) holds. 
Thus, the minimum is obtained at $k=\hat{k}_1$ and $t=R_n^i$. 

($\romannumeral2$)  $H(k,S_n^i,t)$ is a convex function of $k$. Let $\frac{\partial H(k,S_n^i,t)}{\partial k}=0$,
we  get $\hat{k}_2$ (see (\ref{Corollary_1_eq_32})) and 
$
\hat{k}_2\in[-1,1]
$
is equivalent to (\ref{Corollary_1_eq_41}). 
From $S_n^i\leq\hat{s}_n^i$ and $\alpha>0$,  we get  $0\leq(1+\frac{S_n^i}{\hat{s}_n^i})^{1/(\alpha+2)}\leq2$ which is equivalent to  $\bigg|(1+\frac{S_n^i}{\hat{s}_n^i})^{1/(\alpha+2)}-1\bigg|\leq1$. Since $t\geq\tau^{+}\geq\tau$, the inequality (\ref{Corollary_1_eq_41}) holds. 
Thus, the minimum is obtained at $k=\hat{k}_2$ and $t=R_n^i$.

Thus, from ($\romannumeral1$) and ($\romannumeral2$), we obtain
\begin{align}\label{Corollary_1_eq_42_c}
\begin{split}
&\min_{k\in[-1,1], \Delta s\in[-S_n^i,S_n^i],t\in[\tau^{+},R_n^i]}H(k,\Delta s,t)\\
&=\min\{H(\hat{k}_1,-S_n^i,R_n^i),H(\hat{k}_2,S_n^i,R_n^i)\}.\ 
\end{split}
\end{align}	
Moreover, we have the following result:
\begin{align}
\begin{split}
&\min_{k\in[-1,1], \Delta s\in[-S_n^i,S_n^i],t\in[\tau_l,R]}H(k,\Delta s,t)\\
&=\lim_{\bar{\tau}_{l}\rightarrow(\tau_l)^+}\min_{k\in[-1,1], \Delta s\in[-S_n^i,S_n^i],t\in[\bar{\tau}_l,R]}H(k,\Delta s,t)\\
&=\lim_{\bar{\tau}_{l}\rightarrow(\tau_l)^+}\min\{H(\hat{k}_1,-S_n^i,R),H(\hat{k}_2,S_n^i,R)\}.
\end{split}
\end{align}
Denote 
\begin{align*}
\hat{k}_1^+&=\frac{\bigg((1-\frac{S_n^i}{\hat{s}_n^i})^{1/(\alpha/2+1)}-1\bigg)\tau_l^2-(R_n^i)^2}{2R_n^i\tau_l}, \\
\hat{k}_2^+&=\frac{\bigg((1+\frac{S_n^i}{\hat{s}_n^i})^{1/(\alpha/2+1)}-1\bigg)\tau_l^2-(R_n^i)^2}{2R_n^i\tau_l}.
\end{align*}
Thus,  the minimum of $H(k,\Delta s,t)$  for the case of  ($\uppercase\expandafter{\romannumeral2}$) is
\begin{align}\label{result_2_c}
\begin{split}
&\min_{k\in[-1,1], \Delta s\in[-S_n^i,S_n^i],t\in[\tau_l,R_n^i]}H(k,\Delta s,t)\\
&=\min\{H(\hat{k}_1^+,-S_n^i,R_n^i),H(\hat{k}_2^+,S_n^i,R_n^i)\}\\
&\leq\min\{ H(\hat{k}_1^-,-S_n^i,\tau_l), H(\hat{k}_2^-,S_n^i,\tau_l)\}.
\end{split}
\end{align}
Therefore, from
(\ref{result_1_c}) and (\ref{result_2_c}), we obtain
\begin{align}
\begin{split}\label{mfq-53}
&\min_{k\in[-1,1], \Delta s\in[-S_n^i,S_n^i],t\in[0,R_n^i]}H(k,\Delta s,t)\\
&=\min\bigg\{H(-1,-S_n^i,t_1^-), H(1,S_n^i,t_2^-),\\
&~~~~~~~~~~H(\hat{k}_1^+,-S_n^i,R_n^i),H(\hat{k}_2^+,S_n^i,R_n^i),0\bigg\}.
\end{split}
\end{align}
Base on Equations (\ref{Corollary_1_eq_2}) and (\ref{mfq-53}), we  complete the proof.\ ~~~ $\Box$

\section{Proof of Proposition \ref{proposition_2}}
Note that the current remainder bound is $\mathcal{B}_{1}^i\times\cdots$ $\times\mathcal{B}_{N}^i$. The remainder of $f(\bm{x})=\sum_{n=1}^Nf_n(\bm{x}_n)$ (see (\ref{eqsm_4})) is $\Delta f=\sum_{n=1}^N\Delta f_n$ and
$
\Delta f\in \mathcal{B}^{i}$  as shown in (\ref{eqsm_12}).

$\bm{x}_{j}\in\mathcal{E}_{j}^i$ is equivalent to $\bm{x}_{j}=\hat{\bm{x}}_{j}^i+\hat{\bm{E}}_{j}^i\bm{u}_{j},$ $\hat{\bm{x}}_{j}^i=[\hat{s}_{j}^i,(\hat{\rho}_{j}^i)^T]^T$,  $\hat{\bm{E}}_{j}^i=\begin{bmatrix}
S_{j}^i&0\\
0&\bm{E}_{j}^i
\end{bmatrix}$, $\bm{E}_{j}^i$ is a Cholesky factorization of $\bm{P}_{j}^i$, $\bm{u}_{j}=[u_{1,j},u_{2,j}^T]^T$, $|u_{1,j}|\leq1$, $\parallel u_{2,j}\parallel\leq1$,  
$j=1,\cdots,N$, then
\begin{align}
\label{therom_eqsm_16}
s_{n}-\hat{s}_{n}^{i+1}&=\hat{s}_{n}^i+S_{n}^iu_{1,j}-\hat{s}_{n}^{i+1},
\\
\label{therom_eqsm_16_2}
\rho_{n}-\hat{\rho}_{n}^{i+1}&=\hat{\rho}_{n}^i+\bm{E}_{n}^iu_{2,j}-\hat{\rho}_{n}^{i+1}.
\end{align}

Denote $\bm{y}^+$, $f^+$ and $\bm{\varepsilon}^+$ as shown in Proposition \ref{proposition_2}.
We have
\begin{align}
\begin{split}\label{therom_eqsm_17}
\bm{y}^{+}&=f^{+}(\bm{x})+\bm{\varepsilon}^+\\
&=f^+(\hat{\bm{x}}^{i})+\bm{J}^{+,(i+1)}\hat{\bm{E}}^{i}\bm{u}+\hat{\bm{e}}^{+,i}+\bm{e}^{\varepsilon,+}\\
&+\text{diag}(\frac{\hat{\bm{b}}^{+,i}+\bm{b}^{\varepsilon,+}}{2})\Delta^{+}.
\end{split}
\end{align}
Moreover, 
\begin{align}
\begin{split}\label{therom_eqsm_17_2}
\bm{y}_{l}&=f_{l}(\bm{x})+\bm{\varepsilon}_{l}\\
&=f_{l}(\hat{\bm{x}}^{i})+\bm{J}^{l,(i+1)}\hat{\bm{E}}^{i}\bm{u}+\bm{e}^{\varepsilon}_l\\
&-\frac{b^{\varepsilon}_l}{2}+D^{f,i}_l+\Delta_{l},
\end{split}
\end{align}
where $l\in\mathcal{L}^{-,i}$ $\bm{u}=[\bm{u}_{1}^T,\cdots,\bm{u}_{N}^T]^T$, $n=1,\cdots,N$, $\Delta_{y_{l}}\geq0$.

If we denote
$\xi=[1,\bm{u}^T,(\Delta^{+})^T]^T,$ and
 $\Phi_{n}^{s,(i+1)}$, 
$\Phi_{n}^{\rho,(i+1)}$, 
$\Psi^{+,(i+1)}$, 
$\Psi_{l}^{-,(i+1)}$ are shown in (\ref{mfq-34})-(\ref{mfq-36}),  
then (\ref{therom_eqsm_16}), (\ref{therom_eqsm_17_2}) can be written as
\begin{align}
s_{n}-\hat{s}_{n}^{i+1}&=\Phi_{n}^{s,(i+1)}\xi,\label{therom_eqsm_18}\\
\rho_{n}-\hat{\rho}_{n}^{i+1}&=\Phi_{n}^{\rho,i+1}\xi,\label{therom_eqsm_18_2}\\
0&=\Psi^{+,(i+1)}\xi,\label{therom_eqsm_19}\\
\xi^T\Psi_{l}^{-,(i+1)}\xi&\leq0, l\in\mathcal{L}^{-,i}.\label{therom_eqsm_19_2}
\end{align}

Moreover, using (\ref{therom_eqsm_18})-(\ref{therom_eqsm_19_2}) and $\bm{x}_j=\hat{\bm{x}}_{j}^i+\hat{\bm{E}}_{j}^i\bm{u}_{j},\ \ \bm{u}_{j}=[u_{1,j},u_{2,j}^T]^T,\ |u_{1,j}|\leq1,\ \parallel u_{2,j}\parallel\leq1,j=1,\cdots,N$, the conditions that $\bm{x}_j\in\mathcal{E}_{j}^i$ and $\Delta f_j(\bm{u}_j)\in\mathcal{B}_{j}^i$ are relaxed to
\begin{align}
&(\Phi_{n}^{s,(i+1)}\xi)^T(S_{n}^{i+1})^{-2}(\Phi_{n}^{s,(i+1)})\xi)\leq1,\label{therom_eqsm_22}
\end{align}
whenever
\begin{align}
|\bm{u}_{1,j}|&\leq1,\label{therom_eqsm_23}\\
\parallel\bm{u}_{2,j}\parallel&\leq1,j=1,\cdots,N,\label{therom_eqsm_23_2}\\
|\Delta^{+}_l|&\leq1,l\in\mathcal{L}^{+,i},\label{therom_eqsm_24}\\
\xi^T((\Psi^{+,(i+1)})^T\Psi^{+,(i+1)}\xi&=0,\label{therom_eqsm_26}
\\
\xi^T\Psi_{l}^{-,(i+1)}\xi&\leq0, l\in\mathcal{L}^{-,i}\label{therom_eqsm_26_2},
\end{align}
and
\begin{align}
&(\Phi_{n}^{\rho,(i+1)}\xi)^T(\bm{P}_{n}^{i+1})^{-1}(\Phi_{n}^{\rho,(i+1)})\xi)\leq1,\label{therom_eqsm_22_2}
\end{align}
whenever (\ref{therom_eqsm_23})-(\ref{therom_eqsm_26_2}) are satisfied.
The equations (\ref{therom_eqsm_23})-(\ref{therom_eqsm_26_2}) are equivalent to
\begin{align}
\xi^T\text{diag}(-1,diag(\overbrace{0,\cdots,\bm{I}_1,\cdots,0}^{the\ j-th\ entry\ is\ \bm{I}_1}),0)\xi&\leq0,\label{therom_eqsm_23-t}\\
\xi^T\text{diag}(-1,diag(\overbrace{0,\cdots,\bm{I}_2,\cdots,0}^{the\ j-th\ entry\ is\ \bm{I}_2}),0)\xi&\leq0,\label{therom_eqsm_23_2-t}\\
\xi^T diag(-1,\vdots 0,\cdots,0,\overbrace{ 0,\cdots1,\cdots,0}^{the\ l-th\ entry\ is\ 1})\xi\leq&0,\label{therom_eqsm_25-t}\\
\xi^T((\Psi^{+,(i+1)})^T\Psi^{+,(i+1)}\xi=&0,\label{therom_eqsm_26-t}\\
\xi^T\Psi_{l}^{-,(i+1)}\xi\leq0, l\in\mathcal{L}^{-,i}&.
\end{align}
By S-procedure, a sufficient condition such that the inequalities (\ref{therom_eqsm_23})-(\ref{therom_eqsm_26}) imply (\ref{therom_eqsm_22}) to hold is that there exist scalars $\tau^{y_+}$ and nonnegative scalars $\tau_{j}^{1}\geq0$,  $\tau_{j}^{2}\geq0$, $j= 1,\cdots, N,$  $\tau_{l}^{+}\geq0$, $l\in\mathcal{L}^{+,i}$, $\tau_{l}^{-}\geq0$, $l\in\mathcal{L}^{-,i}$,
such that
\begin{align*}
\begin{split}
&(\Phi_{n}^{s,(i+1)})^T(S_{n}^{i+1})^{-2}(\Phi_{n}^{s,(i+1)})-diag(1,0,0)\\
&-\sum_{j=1}^N\tau_{j}^{1} diag(-1,diag(\overbrace{0,\cdots,\bm{I}_1,\cdots,0}^{the\ j-th\ entry\ is\ \bm{I}_1}),0)\\
&-\sum_{j=1}^N\tau_{j}^{2} diag(-1,diag(\overbrace{0,\cdots,\bm{I}_2,\cdots,0}^{the\ j-th\ entry\ is\ \bm{I}_2}),0)\\
&-\sum_{l\in\mathcal{L}^{+,i}}\tau_{l}^{+} diag(-1,\vdots 0,\cdots,0,\vdots\overbrace{ 0,\cdots1,\cdots,0}^{the\ l-th\ entry\ is\ 1})\\
&-\sum_{l\in\mathcal{L}^{-,i}}\tau_{l}^{-}\Psi_{l}^{-,(i+1)}-\tau^{y_+} ((\Psi^{+,(i+1)})^T\Psi^{+,(i+1)})\preceq0.
\end{split}
\end{align*}

The following proof is similar to the proof of Theorem 3.1 in \cite{Shen-Zhu-Song-Luo11}.  Thus, inequalities (\ref{therom_eqsm_22})-(\ref{therom_eqsm_26_2}) are
equivalent to (\ref{therom_eqsm_35}).

In the same way, from (\ref{therom_eqsm_22_2}) and (\ref{therom_eqsm_23})-(\ref{therom_eqsm_26_2}) , the problem (\ref{therom_eqsm_33_2})-(\ref{therom_eqsm_35_2}) has been obtained.  This completes the proof.  ~~$\Box$
\section{Proof of Lemma \ref{lemma-1}}
The problem of finding the lower bound $\breve{D}_{n,l}^i$  of the function $f_{n,l}$ is equivalent to 
\begin{align}
\begin{split}\label{proof-lemma-1}
&\min \frac{s_n}{\|\rho_n-r_l\|^{\alpha}}\\
\text{subject\ to\ } &s_n\in\mathcal{E}_{n}^{s,i},~  \rho_{n}\in\mathcal{E}_{n}^{\rho,i},~\alpha\in[\alpha_{1},\alpha_{2}].
\end{split}
\end{align}
Since  $\frac{s_n}{\|\rho_n-r_l\|^{\alpha}}$ is a monotone function of $s_n$ and $\alpha$, 
the problem (\ref{proof-lemma-1}) has the same optimal value with the following problem 
\begin{align}
\begin{split}\label{proof-lemma-3}
&\min \frac{\hat{s}_n^i-S_n^i}{\max\{\|\rho_n-r_l\|^{\alpha_1},\|\rho_n-r_l\|^{\alpha_2}\}}\\
 &\text{subject\ to\ }  \rho_{n}\in\mathcal{E}_{n}^{\rho,i}.
\end{split}
\end{align}
In order to solve the problem (\ref{proof-lemma-3}), we only need to solve the following problem
\begin{align}
\begin{split}\label{proof-lemma-4}
&\max \|\rho_n-r_l\|\\
&\text{subject\ to\ }  \rho_{n}\in\mathcal{E}_{n}^{\rho,i}.
\end{split}
\end{align}
Similarly, we can  obtain the upper bound $\breve{U}^i_{n,l}$ of the function $f_{n,l}$. This proves the lemma.  $\Box$
\section{Proof of Proposition \ref{proposition_3}}
 Since the function $f_{l,n}(\bm{x}_n)$ is bounded by an interval $[\breve{D}_{l,n},\breve{U}_{l,n}]$, for the $n$-th source, the measurement function is relaxed to 
\begin{align}
y_l=f_{l,n}(\bm{x}_n)+\hat{\varepsilon}_l,~l=1,\cdots,L,
\end{align}
where $\hat{\varepsilon}_l\in[\hat{D}_{l,n},\hat{U}_{l,n}]$, $\hat{D}_{l,n}=\sum_{i\neq n}\breve{D}_{l,i}+\bm{D}_{\varepsilon}(l)$, and $\hat{U}_{l,n}=\sum_{i\neq n}\breve{U}_{l,i}+\bm{U}_{\varepsilon}(l)$.
Thus,  for  $~l=1,\cdots,L,$
\begin{align}
\begin{split}\label{mfq-4}
&\bigg(\max\{\frac{y_l-\hat{U}_{l,n}^i}{g_l},0\}\bigg)^{2/\alpha}\leq \frac{s_n^{2/\alpha}}{\parallel\rho_n-r_l\parallel^2}\\
&~~~~~~~~~~~~~~~~~~~~~~~\leq \bigg(\frac{y_l-\hat{D}_{l,n}^i}{g_l}\bigg)^{2/\alpha}.
\end{split}
\end{align}

Since $\alpha\in[\alpha_{1},\alpha_{2}]\subset[2,4]$ is unknown,  (\ref{mfq-4}) is relaxed to the following  inequations,  for  $~l=1,\cdots,L,$
\begin{align}
\begin{split}\label{mfq-5}
&\tilde{D}_{n,l}\leq \frac{s_n^{2/\alpha}}{\parallel\rho_n-r_l\parallel^2}\leq \tilde{U}_{n,l},
\end{split}
\end{align}
where $\tilde{D}_{n,l}$ and $\tilde{U}_{n,l}$ are defined in Proposition \ref{proposition_3}.
For $l=1,\cdots,L$, we have
\begin{align}
\|\rho_n-r_l\|^{2}&\geq\frac{s_n^{2/\alpha}}{\tilde{U}_{n,l}}\label{mfq-8}.
\end{align}
For $l\in\mathcal{L}^{d}$, i.e.,  $\tilde{D}_{n,l}>0$, we get 
\begin{align}\label{mfq-9}
\|\rho_n-r_l\|^{2}&\leq\frac{s_n^{2/\alpha}}{\tilde{D}_{n,l}}.
\end{align}

Denote 
$\xi=[1,s_n^{2/\alpha},\rho_n]^T$. 
 Thus,   (\ref{mfq-8})
is equivalent to
\begin{align}
\xi^T\Phi_{n,l}^u\xi\leq0, 
\end{align}
and (\ref{mfq-9}) is equivalent to
\begin{align}
\xi^T\Phi_{n,l}^d\xi\leq0,
\end{align}
where $\Phi_{n,l}^u$ and $\Phi_{n,l}^d$ are defined in (\ref{mfq-42}) and (\ref{mfq-39}).

For the $n$-th source,  $\bm{x}_n\in\mathcal{E}_{n}^i$, and  $\alpha\in[\alpha_{1},\alpha_{2}]\subset[2,4]$, we can check that 
\begin{align}
\label{mfq-10}&(\rho_n-\hat{\rho}_{n}^i)^T (\bm{P}_{n}^i)^{-1}(\rho_n-\hat{\rho}_{n}^i)\leq1,\\
\label{mfq-11}&s_n^{2/\alpha}\in[D_{n}^s,U_{n}^s],
\end{align}
where $D_{n}^s$ and $U_{n}^s$  are defined in Proposition \ref{proposition_3}.
The equations (\ref{mfq-10})-(\ref{mfq-11}) are equivalent to
\begin{align}
\begin{split}\label{mfq-12}
\xi^T\Phi_{n}^{\rho}\xi\leq0,\ \
\xi^T\Phi_{n}^s\xi\leq0,
\end{split}
\end{align}
where $\Phi_{n}^{\rho}$ and $\Phi_{n}^s$ are defined in (\ref{mfq-40}) and (\ref{mfq-41}).

The equations
\begin{align}
|s_n^{2/\alpha}-\tilde{s}_{n}|^2&\leq \tilde{S}_{n},\\
(\rho_n-\hat{\rho}_{n}^{i+1})^T (\bm{P}_{n}^{i+1})^{-1}(\rho_n-\hat{\rho}_{n}^{i+1})&\leq1,
\end{align}
are equivalent to
\begin{align}
\label{mfq-13}\xi^T((\Psi^{s}_n)^T\tilde{S}_{n}^{-1}\Psi^{s}_n-diag(1,0,0))\xi&\leq0,\\
\label{mfq-14}\xi^T((\Psi_{n}^{\rho})^T(\bm{P}_{n}^{i+1})^{-1}\Psi_{n}^{\rho}-diag(1,0,0))\xi&\leq0,
\end{align}
where $\Psi^{s}_n$ and $\Psi_{n}^{\rho}$ are defined in Proposition \ref{proposition_3}. 

By S-procedure, a sufficient condition such that the inequalities  (\ref{mfq-8}), (\ref{mfq-9}) and (\ref{mfq-12}) imply (\ref{mfq-13}) to hold is that there exist nonnegative scalars $\tau^1\geq0$,  $\tau^2\geq0$,  $\tau_{l}^3\geq0$, $l = 1, \cdots, L$, $\tau_{l}^4\geq0$, $l\in\mathcal{L}^{d}$,
such that
\begin{align}\label{mfq-15}
\begin{split}
&(\Psi^{s}_n)^Ts_{n,\alpha}^{-1}\Psi^{s}_n-diag(1,0,0)-\tau^2\Phi_{n}^s\\
&-\tau^{1}\Phi_{n}^{\rho}-\sum_{l=1}^{L}\tau_{l}^3\Phi_{n,l}^u-\sum_{l\in\mathcal{L}^{d}}\tau_{l}^4\Phi_{n,l}^d\preceq0.
\end{split}
\end{align}


The following proof is similar to the proof of Theorem 3.1 in \cite{Shen-Zhu-Song-Luo11}.  Thus, inequality (\ref{mfq-15}) is 
equivalent to (\ref{mfq-20}).

In the same way, a sufficient condition such that the inequalities (\ref{mfq-8}), (\ref{mfq-9}) and (\ref{mfq-12}) imply (\ref{mfq-14}) to hold is that the Equation (\ref{mfq-2}) holds. This is the desired conclusion. ~~~~~~~~~$\Box$


\end{document}